\documentclass[10pt,conference]{IEEEtran}
\usepackage{cite}
\usepackage{amsmath,amssymb,amsfonts}
\usepackage{algorithmic}
\usepackage{graphicx}
\usepackage{textcomp}
\usepackage{xcolor}
\usepackage[hyphens]{url}
\usepackage{fancyhdr}
\usepackage{hyperref}

\usepackage{color,soul}
\usepackage[noabbrev,capitalize]{cleveref}
\usepackage{mathtools}

\DeclarePairedDelimiter\floor{\lfloor}{\rfloor}
\usepackage{comment}
\usepackage{enumitem}
\usepackage{xspace}
\usepackage{tcolorbox}
\usepackage{xcolor}
\usepackage{booktabs}
\usepackage{balance}
\hypersetup{
    colorlinks,
    linkcolor={red!50!black},
    citecolor={blue},
    urlcolor={blue!80!black}
}

\newcommand{\hpcayear}{2025}

\newcommand{\red}[1]{{\color{black}#1}}
\newcommand{\purple}[1]{{\color{black}#1}}
\newcommand{\defense}{QPRAC\xspace}
\newcommand{\ignore}[1]{}
\newcommand{\hpcasubmissionnumber}{1419}
\title{\defense{}: Towards Secure and Practical PRAC-based Rowhammer Mitigation using Priority Queues\vspace{-0.1in}}
\def\hpcacameraready{} % Uncomment to build camera-ready version

\newcommand\hpcaauthors{Jeonghyun Woo$^{\dagger, \mathsection}$, Chris S. Lin*, Prashant J. Nair$^\dagger$, Aamer Jaleel$^\ddagger$, Gururaj Saileshwar*}
\newcommand\hpcaaffiliation{$^\dagger$University of British Columbia, $^\ddagger$NVIDIA, *University of Toronto}
\newcommand\hpcaemail{jhwoo36@ece.ubc.ca, shaopenglin@cs.toronto.edu, prashantnair@ece.ubc.ca, ajaleel@nvidia.com, gururaj@cs.toronto.edu\vspace{-0.2in}}

\def\aeopen{}           % The artifact is publically available
\def\aereviewed{}     % The artefact has been reviewed
\def\aereproduced{} % The results have been reproduced
\author{
  \ifdefined\hpcacameraready
    \IEEEauthorblockN{\hpcaauthors{}}
      \IEEEauthorblockA{
        \hpcaaffiliation{} \\
        \hpcaemail{}
      }
  \else
    \IEEEauthorblockN{\normalsize{HPCA \hpcayear{} Submission
      \textbf{\#\hpcasubmissionnumber{}}} \\
      \IEEEauthorblockA{
        Confidential Draft \\
        Do NOT Distribute!!\vspace{-0.25in}
      }
    }
  \fi 
}

\fancypagestyle{camerareadyfirstpage}{%
  \fancyhead{}
  
  \fancyhead[C]{
    \ifdefined\aeopen
    \parbox[][12mm][t]{13.5cm}{\hpcayear{} IEEE International Symposium on High-Performance Computer Architecture (HPCA)}    
    \else
      \ifdefined\aereviewed
      \parbox[][12mm][t]{13.5cm}{\hpcayear{} IEEE International Symposium on High-Performance Computer Architecture (HPCA)}
      \else
      \ifdefined\aereproduced
      \parbox[][12mm][t]{13.5cm}{\hpcayear{} IEEE International Symposium on High-Performance Computer Architecture (HPCA)}
      \else
      \parbox[][0mm][t]{13.5cm}{\hpcayear{} IEEE International Symposium on High-Performance Computer Architecture (HPCA)}
    \fi 
    \fi 
    \fi 
    \ifdefined\aeopen 
      \includegraphics[width=12mm,height=12mm]{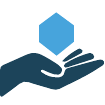}
    \fi 
    \ifdefined\aereviewed
      \includegraphics[width=12mm,height=12mm]{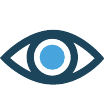}
    \fi 
    \ifdefined\aereproduced
      \includegraphics[width=12mm,height=12mm]{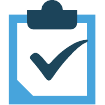}
    \fi
  }
  \fancyfoot[C]{}
}
\begin{document}
\maketitle

%Enables the camera ready header and footer
\ifdefined\hpcacameraready 
  \thispagestyle{camerareadyfirstpage}
  \pagestyle{plain}
\else
  \thispagestyle{plain}
  \pagestyle{plain}
\fi

\newcommand{\hpcaheight}{0mm}
\ifdefined\eaopen
\renewcommand{\hpcaheight}{12mm}
\fi

%%%% PAPER DEFINES %%%%%

\newcommand{\NPRO}{$\text{N}_{\text{PRO}}$\xspace}
\newcommand{\TRH}{$\text{T}_{\text{RH}}$\xspace}
\newcommand{\NBO}{$\text{N}_{\text{BO}}$\xspace}
\newcommand{\ABOACT}{$\text{ABO}_{\text{ACT}}$\xspace}
\newcommand{\ABODELAY}{$\text{ABO}_{\text{Delay}}$\xspace}
\newcommand{\NMIT}{$\text{N}_{\text{mit}}$\xspace}
\newcommand{\NONLINE}{$\text{N}_{\text{online}}$\xspace}
\newcommand{\R}[1]{\text{R}\textsubscript{#1}\xspace}
\newcommand{\RFMAB}{$\text{RFM}_{\text{ab}}$\xspace}
\newcommand{\RFMPB}{$\text{RFM}_{\text{pb}}$\xspace}
\newcommand{\RFMSB}{$\text{RFM}_{\text{sb}}$\xspace}
\newcommand{\RFMSBPB}{$\text{RFM}_{\text{sb/pb}}$\xspace}
\newcommand{\BR}{\text{BR}}
\newcommand{\ALERT}{\text{Alert}}
\newcommand{\RFM}{\text{RFM}}
\newcommand{\ACT}{\text{ACT}}
\newcommand{\PRE}{\text{PRE}}
\newcommand{\TRC}{\text{tRC}}
\newcommand{\TREFW}{\text{tREFW}}
\newcommand{\TREFI}{\text{tREFI}}
\newcommand{\TRFC}{\text{tRFC}}
\newcommand{\REF}{\text{REF}}
\newcommand{\Attackone}{\textit{Toggle+Forget Attack}}
\newcommand{\Attacktwo}{\textit{Fill+Escape Attack}}

%%%%%%%%%%%%%%%%%%%%%%%%%%%%%%%%%%%%%%%%
%%%%%%%% -- PAPER CONTENT STARTS -- %%%%%%%%%

%\input{outline}
\begin{abstract}
JEDEC has introduced the Per Row Activation Counting (PRAC) framework for DDR5 and future DRAMs to enable precise counting of DRAM row activations. PRAC enables a holistic mitigation of Rowhammer attacks even at ultra-low Rowhammer thresholds. PRAC uses an \ALERT{} Back-Off (ABO) protocol to request the memory controller to issue Rowhammer mitigation requests. However, recent PRAC implementations are either insecure or impractical. For example, Panopticon, the inspiration for PRAC, is rendered insecure if implemented per JEDEC's PRAC specification. On the other hand, the recent UPRAC proposal is impractical since it needs oracular knowledge of the `top-N' activated DRAM rows that require mitigation.

This paper provides the first secure, scalable, and practical RowHammer solution using the PRAC framework. The crux of our proposal is the design of a priority-based service queue (PSQ) for mitigations that prioritizes pending mitigations based on activation counts to avoid the security risks of prior solutions. This provides principled security using the reactive ABO protocol. Furthermore, we co-design our PSQ, with opportunistic mitigation on Refresh Management (\RFM{}) operations and proactive mitigation during refresh (\REF{}), to limit the performance impact of ABO-based mitigations. \defense{} provides secure and practical RowHammer mitigation that scales to Rowhammer thresholds as low as 71 while incurring a 0.8\% slowdown for benign workloads, which further reduces to 0\% with proactive mitigations.
\end{abstract}
\section{Introduction}
Relentless scaling of Dynamic Random Access Memory (DRAM) technology has exposed critical security vulnerabilities like Rowhammer (RH). RH exploits inter-cell interference to rapidly activate DRAM rows, causing bit-flips in neighboring victim rows~\cite{seaborn2015exploiting, frigo2020trrespass, gruss2018another, aweke2016anvil, cojocar2019eccploit, gruss2016rhjs, vanderveen2016drammer}. The number of activations needed to induce bit-flips, known as the Rowhammer threshold (\TRH), has dropped from 70K~\cite{kim2014flipping} to 4.8K~\cite{kim2020revisitingRH} and is expected to decrease further with each generation. To counteract RH, the DRAM industry has proposed a series of in-DRAM mitigations, with the latest being Per Row Activation Counting (PRAC)~\cite{jedec_ddr5_prac}. However, the PRAC specification provides minimal implementation details, placing significant responsibility on DRAM manufacturers. This paper introduces a solution to implement PRAC securely and practically in DRAM for ultra-low \TRH values (sub-100).

\begingroup
\renewcommand\thefootnote{$\mathsection$}%
\footnotetext{A large part of this work was performed while Jeonghyun Woo was interning with NVIDIA Research.}
\endgroup

Prior in-DRAM RH mitigations implemented by DRAM vendors commercially have repeatedly fallen short in either security or scalability. For example, DDR4 devices use Targeted Row Refresh (TRR), which relies on a tracker to identify aggressor rows and refresh neighboring victim rows~\cite{hassan2021UTRR}. However, these trackers can only monitor a limited number of rows and are vulnerable to attack patterns like TRRespass, which target a larger number of rows~\cite{frigo2020trrespass}. DDR5 introduced the Refresh Management (\RFM{}) command to mitigate victim rows proactively. This limits the number of activations per bank before an \RFM{}-based mitigation needs to be issued. However, such solutions do not scale to \TRH below 100. Even state-of-the-art defenses like PrIDE~\cite{jaleel2024pride} and MINT~\cite{MINT} require frequent \RFM{}s (e.g., 1 \RFM{} every 10 activations), resulting in nearly 30\% activation bandwidth loss at \TRH of 250. Consequently, JEDEC, the DRAM standards committee, proposed PRAC for DDR5 DRAM chips (and beyond)~\cite{jedec_ddr5_prac}.

PRAC maintains activation counters for each row in DRAM and allows the DRAM to use the \ALERT{} Back-Off (ABO) protocol to request an \RFM{} from the host only when mitigation is needed. The ABO protocol uses the \textit{Alert\_n} pin in the DRAM module to notify the memory controller when any row activation exceeds the Back-Off threshold (\NBO{}), which is set lower than the \TRH. This prompts the memory controller to issue \RFM{} commands on demand and perform RH mitigation before a row reaches \TRH. Although recent work explored this approach for ultra-low \TRH (sub-100), they face security concerns or impractical overheads.

\begin{figure*}[!htb]
    \centering
\includegraphics[width=6.7in]{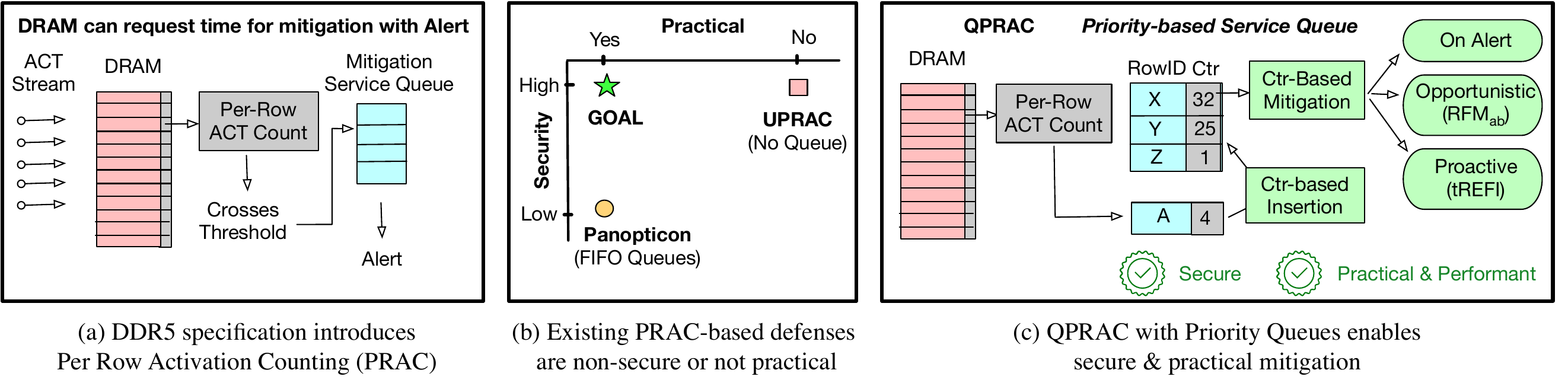}
    \vspace{-0.1in}
    \caption{(a) With the PRAC framework, DRAM can request a time for mitigation when it needs it (based on per-row activation counters), using \ALERT{}s to service its mitigation queue. (b) Existing PRAC implementations are either insecure (Panopticon~\cite{bennett2021panopticon}) due to the usage of FIFO-based queues or impractical (UPRAC~\cite{UPRAC}) due to the lack of any queues. (c) We propose \defense{}, using a priority-based service queue (PSQ) for mitigations, which can be cleared on \ALERT{}s but also \textit{opportunistically} when another bank requests an All-Bank \RFM{} or \textit{proactively} on \REF{}s. We design \defense{} to be both secure and practical.
    }
    \vspace{-0.2in}
    \label{fig:intro}
\end{figure*}

\smallskip 
\noindent \textbf{1. Lack of Security}: Panopticon~\cite{bennett2021panopticon}, the inspiration behind PRAC, also uses in-DRAM per-row activation counters and a FIFO-based service queue to track rows exceeding an activation threshold. When the queue is full, the DRAM module uses the ABO protocol to stop activations and request an \RFM{} for RH mitigation, thus freeing up space in the queue.

However, Panopticon is insecure under the PRAC specification. PRAC employs a non-blocking ABO protocol, allowing the memory controller to continue issuing additional activations for up to 180ns. This window permits a row to surpass the mitigation threshold while bypassing the service queue if the queue is already full. Moreover, Panopticon only selects a row for mitigation when the threshold bit (t) in its counter toggles. Thus, the next insertion for a bypassed row occurs only after $2^{t}$ activations. We show that this leads to unmitigated rows being activated up to 50$\times$ higher than \TRH, compromising security.

\smallskip 

\noindent \textbf{2. Impractical Overheads}: UPRAC~\cite{UPRAC} proposes a PRAC implementation \emph{without} a service queue. This design triggers an \ALERT{} when any DRAM row exceeds \NBO{}. Thereafter, it mitigates the top-N activated rows with N subsequent \RFM{}s.

While this avoids the security issues of service queues, it incurs impractical performance overheads. During each \ALERT{}, a bank must search through activation counters for \emph{all} rows to identify the top-N rows, which is impractical. For example, in 32Gb DRAM chips with 128K rows per bank, identifying the top-N rows requires activating (52ns) and reading PRAC counters from each row. This causes an impractical overhead, locking all the banks for multiple milliseconds per \ALERT{}.

\smallskip
\noindent \textbf{Our Proposal}: We propose \defense{}, the first secure and practical PRAC implementation adhering to the JEDEC standard to address these concerns. \defense{} uses a low-cost and practical service queue to mitigate the highest activated rows while providing security bounds for ultra-low \TRH values.

\noindent \textbf{1. Priority-Based Service Queue}: \defense{} uses a small Priority-Based Service Queue (PSQ) to track frequently activated rows (see Figure~\ref{fig:intro}(c)). Each PSQ entry maintains a row address and its activation count, using the activation count as the priority. While the PSQ can only track a few rows, it is PRAC-aware, enabling precise tracking of activation counts for all DRAM rows. On activations, the PSQ compares the in-DRAM row activation count with existing PSQ entries. If the activated row has a higher count than any PSQ entry, the entry with the lowest count is replaced with the activated row and its count. Thus, using an N-entry queue, the PSQ maintains the top-N highest activated rows between mitigations. As such, the PSQ avoids the security pitfalls of Panopticon, where the FIFO-based service queue can be bypassed when full.

We analytically show that the security of \defense{} with a PSQ is identical to an ideal PRAC implementation if the PSQ size is at least the number of \RFM{}s per \ALERT{} (1, 2, or 4). We craft optimized versions of the wave\cite{UPRAC} or feinting attacks\cite{ProTRR}, incorporating the effects of transitive attack mitigations in PRAC, and show \defense{} is secure up to \TRH of 44, 29, and 22 for 1, 2, or 4 \RFM{}s~/~\ALERT{} with an \NBO{} of 1, respectively.

\smallskip 
\noindent \textbf{2. Opportunistic Mitigation}: At ultra-low \TRH (sub-100), \defense{} frequently triggers \ALERT{}s due to low \NBO{} values. The ABO protocol issues All-Bank \RFM{}s as the current DRAM interface cannot identify the specific bank that initiated the \ALERT{}. Consequently, all banks receive \RFM{} commands simultaneously, enabling other banks to \textit{opportunistically} mitigate rows, even if their activation counts are below \NBO{}. This avoids future \ALERT{}s for these rows and improves performance.
\purple{For instance, at \NBO{} of 32, QPRAC without opportunistic mitigations incurs a significant 12.4\% overhead. Opportuinsitc mitigations considerably reduce this overhead to just 0.8\%.}

\noindent \textbf{3. Proactive Mitigation}: 
DRAM devices receive periodic refresh commands (\REF{}) to ensure charge retention in DRAM cells. Existing DDR4 and DDR5 DRAM currently issue RH mitigations\cite{hassan2021UTRR, ProTRR, jattke2024zenhammer} in the shadow of such \REF{} commands. QPRAC can also leverage these mitigations by \textit{proactively} mitigating the highest activated row in the PSQ during a \REF{}. These mitigations provide modest security benefits, reducing \TRH{} by 4 to 10 activations at \TRH below 100, and offer performance benefits by lowering \ALERT{} frequency. For example, at \NBO{} of 32, QPRAC with proactive mitigations reduces the slowdown from 0.8\% to 0\%. 
However, performing proactive mitigation on every \REF{} has high energy overheads, increasing energy consumed by 14.6\%. 
Moreover, not all proactive mitigations are useful, as many entries from the PSQ may not even reach \NBO{}. Leveraging this, we propose an energy-optimized design that performs proactive mitigation \emph{only} when the activation count of the highest activated row in the PSQ meets or exceeds a Proactive Mitigation threshold (\NPRO{}). This reduces the energy overhead of proactive mitigation from 14.6\% to 1.9\%,
without impacting performance.

\smallskip
\noindent Overall, this paper makes five key contributions:
\begin{enumerate}[leftmargin=*]
    \item We show that existing PRAC implementations are either insecure or impractical.
    \item We argue that the service queue design in PRAC is crucial for security, especially with a non-blocking \ALERT{} specification, potentially leading to overwhelmed queues.
    \item We propose the first practical and secure PRAC implementation, \defense{}, using priority queues.
    \item We analyze the security of \defense{} and show that it is secure up to a minimum \TRH as low as 22.
    \item We realize the potential of \textit{opportunistic} mitigation and co-design \defense{} with a \textit{proactive} mitigation mechanism, providing performance and security benefits. 
    \item We further enhance our design, including an energy-optimized proactive mitigation scheme, significantly reducing energy overhead while preserving performance benefits.
\end{enumerate}

\defense{}, with an \NBO{} of 32 and one mitigation per \ALERT{},  securely handles a \TRH{} of 71. Across 57 workloads, 
\defense{} incurs 0.8\% slowdown, which goes down to 0\% with Proactive Mitigation, compared to a non-secure baseline without \ALERT{}s. \defense{} requires $<$15 bytes of storage per DRAM bank. 
\section{Background and Motivation}\label{sec:background}
\subsection{Threat Model}\label{threat_model}
We assume an attacker can issue memory requests for arbitrary rows, knowing the defense algorithm. Our defense aims to prevent single-sided and multi-sided Rowhammer attacks~\cite{frigo2020trrespass,jattke2021blacksmith} and attacks like Half-Double~\cite{HalfDouble}. The RowPress~\cite{rowpress} attack is out of scope since its effects are orthogonal, and it can be mitigated by limiting row open time.

\subsection{The Rowhammer Vulnerability}
Rowhammer (RH) is a read disturbance error where rapid activations of specific rows (aggressors) accelerate charge leakage in neighboring victim rows, leading to bit-flips. The minimum activations required to cause these bit-flips is the RH Threshold (\TRH). As DRAM technology scales, \TRH values have decreased significantly, dropping nearly 16$\times$ from 70K in 2014~\cite{kim2014flipping} to 4.5K in 2020~\cite{kim2020revisitingRH}, with thresholds likely to drop further in future generations. At ultra-low \TRH, RH is a bigger security~\cite{frigo2020trrespass, kwong2020rambleed, HalfDouble, gadgethammer_sec24, jattke2024zenhammer, gruss2018another} and reliability risk~\cite{loughlin2022moesi}. For future DRAM, it is beneficial to have solutions catering to  \TRH of 100 or lower.

\subsection{In-DRAM Rowhammer Mitigation}
\subsubsection{Commercial Solutions} Until recently, the DRAM industry relied on two primary approaches to mitigate Rowhammer:
\begin{itemize}[leftmargin=*]
\item \textbf{Targeted Row Refresh (TRR)}: DRAM manufacturers implemented Targeted Row Refresh (TRR) in DDR4, LPDDR5, and HBM2 to prevent RH\cite{HBM2-characteristic, jattke2021blacksmith}. TRR tracks potential aggressor rows using small counter tables or probabilistically and refreshes neighboring victim rows within a blast radius (BR) around the aggressor (e.g., $\text{BR}=2$ means two victim rows on either side of the aggressor row are refreshed) every few \REF{}s\cite{hassan2021UTRR, ProTRR}. However, due to limited storage in DRAM, these counter tables have only a few entries, making them vulnerable to malicious patterns that thrash the tables and bypass protections. Several attacks\cite{jattke2021blacksmith, ProTRR, frigo2020trrespass, hassan2021UTRR, jattke2024zenhammer} have circumvented TRR and induced RH bit-flips.

\item \textbf{Refresh Management (\RFM{})}: As \REF{}-based mitigations do not scale with decreasing \TRH values, DDR5 introduced the Refresh Management (\RFM{}) command. This command allows the memory controller to track the total activations issued to a bank. When the number of activations exceeds a specified threshold, the controller issues an \RFM{} command, giving the DRAM time to perform mitigations.
\end{itemize}

\subsubsection{Academic Solutions} State-of-the-art in-DRAM solutions, such as PrIDE\cite{jaleel2024pride} and MINT\cite{MINT}, use probabilistic sampling and FIFO-based or single-entry trackers. These solutions can tolerate a \TRH{} of 1700 with one mitigation per \TREFI{} and a \TRH{} of up to 400 with 4 \RFM{}s per \TREFI{} with negligible slowdown. However, scaling to lower \TRH{} (250 or lower) requires additional \RFM{}s. For example, performing an \RFM{} every 10 activations incurs an activation bandwidth loss of nearly 30\%, as each \RFM{} takes 350ns. This results in prohibitive performance overheads. 

\subsection{Per Row Activation Counting (PRAC)}
JEDEC's DDR5 specification\cite{jedec_ddr5_prac} introduces the Per Row Activation Counting (PRAC) framework to precisely count row activations and holistically mitigate Rowhammer attacks. PRAC consists of two key mechanisms: (1) an activation counter added to each DRAM row, using additional DRAM cells and sense amplifiers, which is incremented on each activation of the corresponding row, and (2) the \ALERT{} Back-Off (ABO) protocol, which the DRAM uses to request extra mitigation time when a Rowhammer threat is detected.

The ABO protocol allows DRAM to assert the \ALERT{} signal when a row's activation counter crosses a Back-Off threshold (\NBO{}). 
The memory controller can then issue activations for only 180ns (\ABOACT{} activations) before sending a pre-configured number of \RFM{}s (\NMIT{}) to allow DRAM to perform RH mitigations to victim rows. After these mitigations, the next \ALERT{} can be asserted by the ABO protocol \emph{only} after the DRAM services a pre-specified number of row activations (\ABODELAY{}). Additionally, PRAC requires updated DRAM timings to account for the time needed to increment the in-DRAM activation counters. \cref{table:prac_params} showcases the PRAC-related parameters, their explanation, and values for our evaluation.

\begin{table}[!htb]
  \centering
\vspace{-0.1in}
  \caption{PRAC Parameters as per DDR5 specification~\cite{jedec_ddr5_prac}}
  \vspace{-0.1in}
  \begin{footnotesize}
  \label{table:prac_params}
  \begin{tabular}{lcc}
    \hline
    \textbf{Parameter} & \textbf{Explanation} & \textbf{Value} \\ \hline
    \rule{0pt}{1\normalbaselineskip}\NBO{}     & Back-Off Threshold & \NBO{} $\leq$ \TRH{} \\ %\hline
    \NMIT{}     & Num \RFM{}s on \ALERT{} & 1, 2, or 4 \\ %\hline
    \ABOACT{}  & Max. \ACT{}s from \ALERT{} to \RFM{} & 3 (up to 180ns)  \\ %\hline    
    \ABODELAY{}       & Min. \ACT{}s after \RFM{} to \ALERT{} & Same as \NMIT{} (1,2, or 4) \\ \hline  
  \end{tabular}
  \end{footnotesize}
 \vspace{-0.2in}
\end{table}

\subsection{Existing PRAC Implementations and their Drawbacks}
While the DDR5 specification proposes the PRAC interface, its implementation is left to the DRAM vendors, making the security guarantees of such solutions unclear. Below, we describe two PRAC implementations and their drawbacks. 

\subsubsection{Insecurity of Panopticon}
\label{sec:pano_attacks} 

Panopticon~\cite{bennett2021panopticon} inspired the PRAC design by proposing activation counters for each DRAM row. When a counter crosses the mitigation threshold, a threshold bit $t$ toggles, identifying the row for mitigation. A \emph{FIFO-based} service queue tracks these rows. If the queue is full, the DRAM uses the ABO protocol to halt activations and request \RFM{}s for RH mitigation, freeing up queue entries. Additionally, \REF{} commands can mitigate rows from the service queue, which further frees up entries.

\smallskip
\noindent \textbf{Vulnerability}: Panopticon is insecure when implemented with the PRAC specification because of three reasons:
(1) mitigating rows only upon toggling of the counter's threshold bit (t-bit), (2) a limited capacity of FIFO-based service queue, and (3) PRAC's non-blocking nature of \ALERT{}s.\footnotemark{}
These issues render a PRAC implementation based on Panopticon insecure. We elaborate on these with two illustrative attacks.

\smallskip
\footnotetext{While these vulnerabilities could be avoided by making Alerts immediately blocking, this would violate JEDEC's PRAC specifications, which allow up to 180ns for servicing Alerts. This delay is necessary for the memory controller to receive the \ALERT{} from DRAM, read the Mode Register, and confirm whether it was triggered by the ABO protocol or other issues, such as CRC errors, before initiating the ABO protocol for RH mitigations\cite{jedec_ddr5_prac}.}

\noindent \textbf{(1) \Attackone{} -- Exploiting t-bit Toggling:} Since PRAC uses a non-blocking \ALERT{}, the memory controller can issue up to \ABOACT{} activations even after receiving an \ALERT{}. These activations can be exploited when the service queue is full, enabling a row to bypass the queue if its $t$-bit is toggled by the \ABOACT{}. After a bypass, subsequent activations to that row will not trigger immediate RH mitigation since the $t$-bit does not toggle again until other $2^{t}$ activations. This allows a row to be activated indefinitely without mitigation until the end of \TREFW{} (32ms). This attack is outlined below.

Assuming a service queue size of Q and a mitigation threshold of M, our attack activates Q+1 rows, starting from the same activation count. The rows are activated uniformly in a round-robin manner. After each row receives M-1 activations, the next activation of the first Q rows pushes each of them into the service queue, filling it, and causing the DRAM to raise the \ALERT{}. While the queue is full, using the \ABOACT{}, the Q+1th row (target row) is activated twice, allowing it to escape insertion into the queue and mitigation. After the \ALERT{} is serviced with the \RFM{}s and the queue is no longer full, the Q rows are activated twice to bring them to the target row activations. Then, the previous attack step is repeated on Q+1 rows~\footnote{While Panopticon~\cite{bennett2021panopticon} can initialize rows with random activation counts, the attacker can easily get rows to a known activation count. The paper~\cite{bennett2021panopticon} shows that activating randomly chosen rows can cause the queues to be full and \ALERT{} raised in tens of minutes, bringing the activation count of the last row to a known value (a multiple of the threshold). This can be repeated Q+1 times, to get Q+1 rows with activation counts at a multiple of the threshold.}. This process continues until the \TREFW{} period (32ms).

\begin{figure}[h]
\centering
\vspace{-0.12in}
\includegraphics[width=0.43\textwidth,height=\paperheight,keepaspectratio]{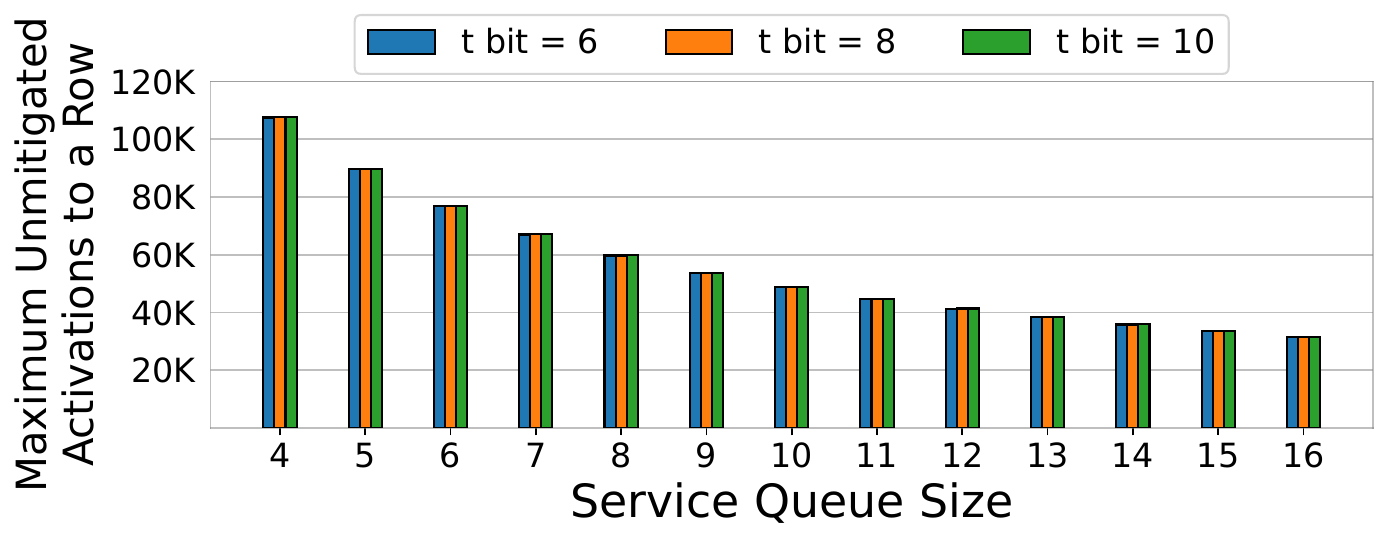}
\vspace{-0.11in}
\caption{The security vulnerability of Panopticon~\cite{bennett2021panopticon} due to t-bit toggling, i.e., maximum activations before the row receives a mitigation with \Attackone. For sub-100 \TRH{}, our attack can cause a DRAM row to receive even 100$\times$\TRH{} activations without any mitigation. This vulnerability is independent of the mitigation threshold ($2^t$) used by Panopticon.}
\vspace{-0.11in}
\label{fig:panopticon_insecure}
\end{figure}

\cref{fig:panopticon_insecure} shows the maximum activations to the target row in the above attack before mitigation, with queue sizes varying from 4 to 16 for different mitigation thresholds of 32, 512, or 1024 (t-bit of 6, 8, or 10 respectively). The target row receives no mitigation until the end of \TREFW{}. Consequently, it can be activated beyond 100K times without mitigation with a queue size of 4 and about 25K times with a queue size of 16. As the queue size increases, the maximum activations decrease linearly as the pool of rows to be uniformly activated increases. No matter the queue size, the target row is activated beyond 100$\times$ of \TRH{} (for \TRH{} of sub-100), compromising the security of Panopticon. 
Moreover, this behavior is independent of the mitigation threshold (t-bit) used, indicating that reducing the mitigation threshold cannot address the vulnerability.

One way to address this vulnerability is to extend Panopticon by using larger counters that never overflow and comparing the full counter value against the threshold to identify rows to be mitigated. Thus, even if \ABOACT{} causes a counter to cross the threshold and temporarily skip mitigation due to the queue being full, the row would be inserted into the queue on subsequent \ACT{}s.
However, this design is still insecure.

\smallskip
\noindent \textbf{(2) \Attacktwo{} -- Exploiting FIFO Service Queues:}\label{sec:attack2} Assuming the full counter-value is compared with the threshold on each activation (so t-bit toggling cannot be exploited), an attacker can still avoid mitigations for a target row by hammering it \textit{only} with \ABOACT{} \textit{and} only when the FIFO-based service queue is full. 

In this attack, the attacker first activates the target row and Q other rows, with M-1 activations each. The attacker then activates Q rows by one activation to make the FIFO queue full. When the \ALERT{} is raised, the three \ABOACT{} activations to a target row can raise the activation count without the row being selected for mitigation. 
After the \ALERT{} is serviced, up to four entries will be removed from the queue (and one extra entry may be removed due to mitigation on \TREFI{}). The attacker will again fill the queue by activating five other rows to M activations. Thus, with every 5~$\times$~M additional \ACT{}s, the target row receives three extra activations (via \ABOACT{}) without a mitigation.

\begin{figure}[h!]
\vspace{-0.12in}
\centering
\includegraphics[width=0.45\textwidth,height=\paperheight,keepaspectratio]{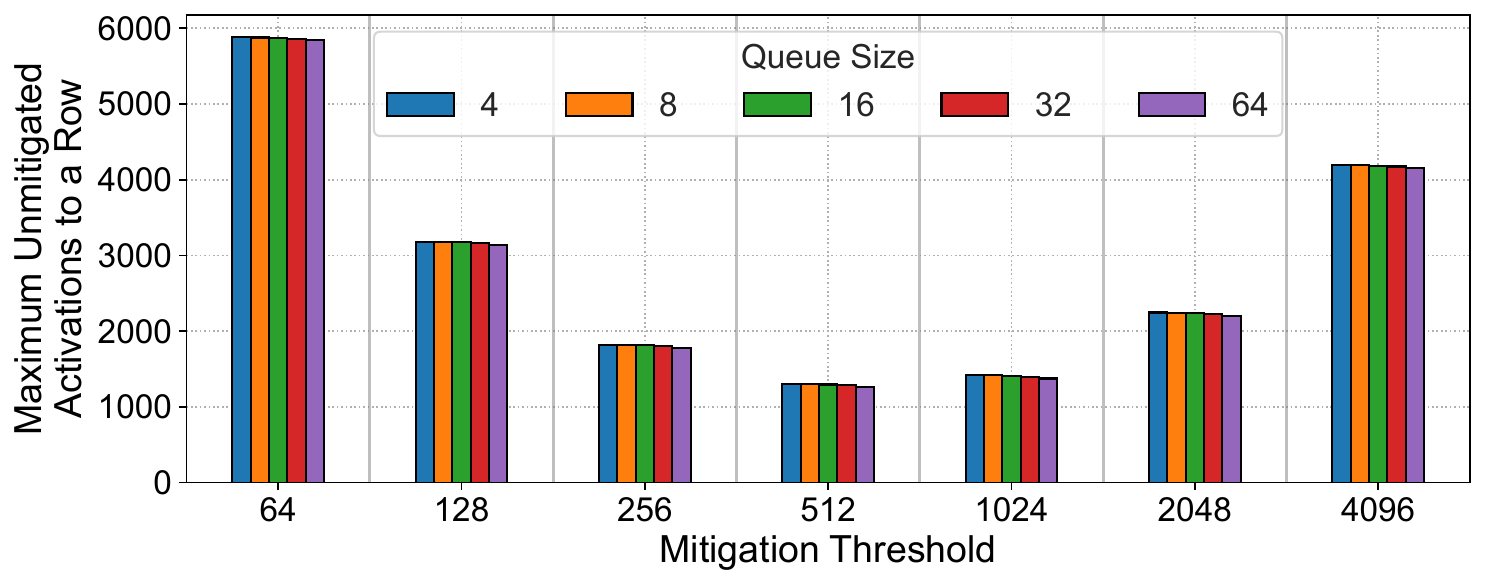}
\vspace{-0.10in}
\caption{The security vulnerability of Panopticon (with full counter comparisons) under the \Attacktwo{}, which exploits filled FIFO-based service queues. Combined with non-blocking \ALERT{}, this allows at least 1283 unmitigated \ACT{}s (at a mitigation threshold of 512), with the number increasing at lower thresholds.
}
\label{fig:uprac+fifo_vulnerability}
\vspace{-0.11in}
\end{figure}

Figure~\ref{fig:uprac+fifo_vulnerability} shows the maximum unmitigated activations to the target row using this attack, as the mitigation threshold (M) ranges from 64 to 4096. The attacker can achieve a minimum of 1283 unmitigated \ACT{}s on the target row at a mitigation threshold of 512. In fact, at lower mitigation thresholds, the number of unmitigated activations to target row increases dramatically, as filling up the FIFO queues requires less effort. 
Thus, even the optimized version of Panopticon is insecure below a \TRH{} of 1280, primarily due to filled FIFO service queues combined with non-blocking \ALERT{}s in PRAC, allowing a high number of unmitigated activations. 

\smallskip

\subsubsection{Impracticality of UPRAC}
UPRAC~\cite{UPRAC} proposes a PRAC implementation \emph{without} a service queue. It raises an \ALERT{} when any DRAM row crosses the Back-Off threshold (\NBO{}) and proposes to mitigate the N highest activated rows globally.

\smallskip
\noindent \textbf{Impractical Overhead of UPRAC:} While this design does not use a service queue and avoids related security vulnerabilities, it is impractical. Without a service queue, on an \ALERT{}, it is impractical for the DRAM to read the activation counters of \emph{all} the DRAM rows to identify the top-N rows.

\noindent \textbf{Vulnerability of UPRAC + FIFO Service Queues:} UPRAC can be practical using a FIFO-based service queue, requiring mitigation when the row activation count exceeds a threshold lower than \NBO{}. However, this approach is also vulnerable to the \Attacktwo{} on Panopticon (\cref{sec:pano_attacks}), which exploit full FIFO queues in combination with non-blocking \ALERT{}s. Entries can be inserted at a maximum rate of one per activation, whereas removal occurs at best at one per four activations (\ABOACT{} + \ABODELAY{}). An attacker can thus fill up the UPRAC FIFO queue with Q rows, each activated to \NBO{}, and then use three \ABOACT{} to hammer the target row each time. This incurs at least 1283 ACTs to a target row without mitigation (at \NBO{} of 512) and higher \ACT{}s at lower \NBO{}. Thus, UPRAC is insecure below the \TRH{} of 1280.
\begin{tcolorbox}[boxsep=1pt,left=4pt,right=4pt,top=2pt,bottom=2pt]
\textbf{Key Question:} Designs without service queues are impractical, and practical designs using FIFO queues are insecure at sub-1000 \TRH{}. Can we design a PRAC \red{implementation} that is both practical and secure at ultra-low \TRH (sub-100)? 
\end{tcolorbox}
\vspace{-0.01in}

\subsection{Goal}
Our goal is to design a secure PRAC implementation that provides strong security guarantees at ultra-low \TRH{} (sub-100). At the same time, we seek a practical service queue design that avoids the insecurity of FIFO-based queues without modifying the JEDEC PRAC specification to ensure that DRAM vendors can easily adopt it. To that end, we explore a priority-based service queue design using activation counts to ensure secure and scalable Rowhammer mitigation.

\section{Design of \defense{}}\label{design}
This paper proposes \defense{}, a PRAC-based implementation to enable practical and secure Rowhammer mitigation. The key focus of our solution is the \textit{Priority-based Service Queue} (PSQ), which provides a practical mechanism to track pending mitigation without the security vulnerabilities introduced by FIFO-based service queues. Below, we provide an overview of \defense{}'s design, its queue management policies (insertion, eviction, and mitigation), and finally, its co-design with other mitigation opportunities available to DRAM.

\subsection{Overview of \defense{}}
Any implementation of the PRAC specification needs to answer the following questions: (1) how to select a row for RH mitigation, (2) how to track rows identified for RH mitigation while it is pending, and (3) when to request the memory controller for additional time for RH mitigation. 

\begin{figure}[t]
 \centering
\includegraphics[width=0.9\textwidth/2,height=4cm,keepaspectratio]
 {"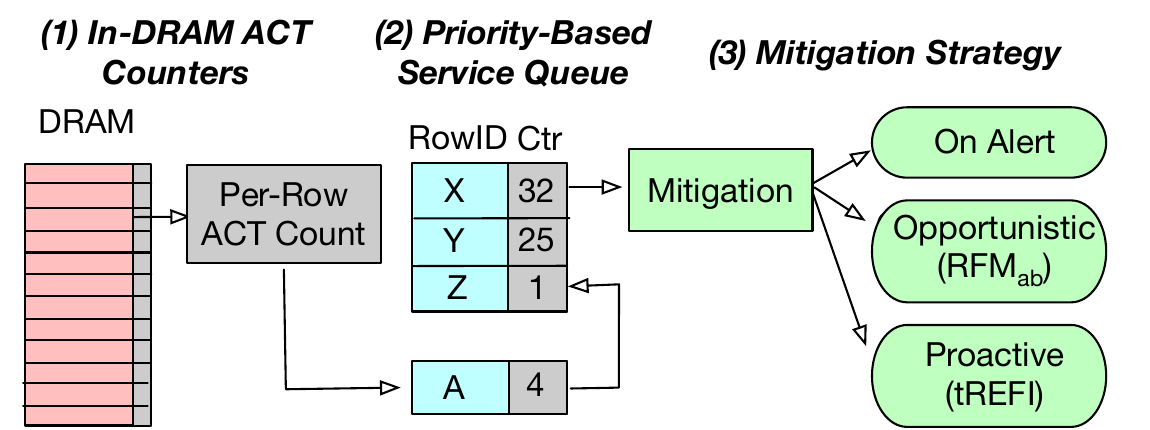"}
 \caption{Overview of \defense{} design. It consists of three components: (1) PRAC-based in-DRAM activation counters, (2) a Priority-Based Service Queue (PSQ) to identify rows to be mitigated, and (3) a strategy that uses \ALERT{}-based, opportunistic, and proactive RH mitigations.}
 \vspace{-0.2in}
 \label{fig:design_overview}
 \end{figure}

As shown in \cref{fig:design_overview}, \defense{} addresses these using three key components: (1) per-row activation counters in DRAM, as specified by PRAC specification~\cite{jedec_ddr5_prac}, (2) a per DRAM bank priority-based service queue (PSQ) that tracks the highest activated rows even in situations where the queue is full, and (3) an implementation of the \ALERT{} Back-Off (ABO) protocol to mitigate rows identified in the PSQ in a timely manner. We now explain these components.

\begin{figure}[b]
 \centering
 \vspace{-0.2in}
 \includegraphics[width=0.9\textwidth/2,height=4cm,keepaspectratio]
 {"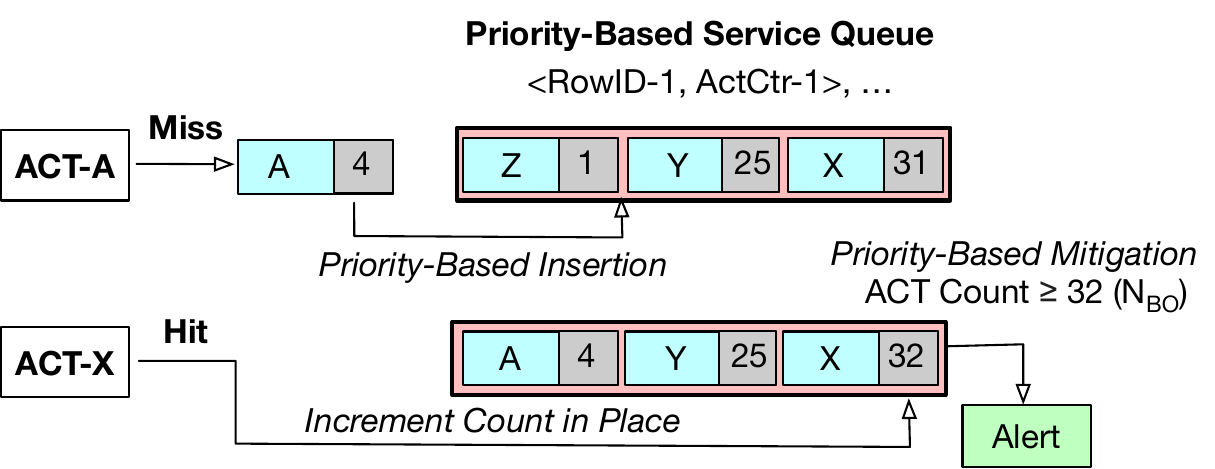"}
 \caption{Design of Priority-Based Service Queue (PSQ). Any activation can insert a row into PSQ based on priority (activation count) on misses and increment count on hits. PSQ raises an \ALERT{} if any count is at \NBO{} or above.}
 \label{fig:psqdesign}
 \end{figure}

\subsection{Priority-Based Service Queue Design and Operation}
\subsubsection{Design} \defense{} consists of an N-entry priority-based service queue (PSQ) per DRAM bank. As shown in \cref{fig:psqdesign}, each entry represents a row and includes its RowID and current activation count. The queue is sorted in descending order by activation count, prioritizing rows with higher activation counts. We design the PSQ using a CAM (content-addressable memory); we assume small PSQs with five entries per bank.

\smallskip
\subsubsection{Operation} When a row is activated, its in-DRAM activation counter is incremented according to the PRAC specification. Simultaneously, the activated row is also considered for insertion into the service queue. The PSQ inserts only rows with activation counts higher than the lowest count in the queue. If a new entry is inserted, the entry with the lowest count is evicted. If the activated row is already in the queue, its activation count is updated to match the in-DRAM count.

\smallskip
\subsubsection{Intuition} Unlike prior works, which are vulnerable when the service queue becomes full, the PSQ is \emph{intentionally} designed to be full at all times. This design ensures that the PSQ \emph{always} retains and tracks the highest activated rows. Consequently, the PSQ cannot lose information about heavily activated rows, even in attack scenarios that activate more rows than the service queue's capacity -- a scenario that compromises the security of previous solutions.

\subsection{Mitigation Policy using \ALERT{} Back-Off Protocol}
\subsubsection{Design} \defense{} uses the ABO protocol to request DRAM mitigations. Unlike prior defenses that use different thresholds for issuing mitigation and signaling an \ALERT{}, \defense{} simplifies this process. It tracks rows with the highest activation counts in the PSQ, using a single threshold to flag the highest priority row for mitigation and raise an \ALERT{}. 

\smallskip
\subsubsection{Operation} When the activation count of the highest activated row in the PSQ crosses the Back-Off threshold (\NBO{}), \defense{} identifies the need for an RH mitigation. It asserts the \ALERT{} signal to the memory controller to initiate the ABO process and issue a mitigation. The memory controller then issues \NMIT{} \RFM{}s (default \NMIT{} is 1) to allow the DRAM to mitigate \NMIT{} rows with the highest counts in the PSQ. For each \RFM{}, the DRAM mitigates the aggressor with the highest activation count in the PSQ, refreshing the blast-radius (BR) victim rows above and below it (default BR is 2). Additionally, the aggressor's in-DRAM per-row counter is reset to 0 by activating it, and its entry is evicted from the PSQ.

To mitigate transitive attacks like Half-Double, each mitigative refresh to the victim row also increments the in-DRAM counter associated with the victim. The victim row itself may be inserted into the PSQ if its activation count is higher than the minimum activation count of entries in the PSQ.

After the \RFM{}s are serviced, the \ALERT{} is de-asserted until \ABODELAY{} activations. The next \ALERT{} can be raised if there are rows in the PSQ with activation counts at or beyond \NBO{}.

\smallskip
\subsubsection{Security} Any row that crosses \NBO{} is eligible for RH mitigation by triggering an \ALERT{}. However, the non-blocking nature of the \ALERT{} allows some activations to occur despite the raised \ALERT{}, and there is a limit on how frequently \ALERT{}s can be raised. This permits a certain number of activations to rows beyond the \NBO{} value. In \cref{sec:security}, we establish an upper bound on these activations to determine the appropriate \NBO{} value for security at a given \TRH{}. Our analysis in \cref{sec:priority_queue_security} shows that as long as the PSQ size matches \NMIT{} (the number of mitigations per \ALERT{}), \defense{} provides deterministic security against RH attacks, even at sub-100 \TRH.

\subsection{Additional Mitigation Opportunities using PSQ}
Thus far, we have considered RH mitigation for a row in the PSQ only when its activation count crosses the Back-Off threshold (\NBO{}), triggering an \ALERT{}. However, there are additional opportunities to provide RH mitigation to entries in the PSQ at no extra performance cost. These additional RH mitigations can reduce the number of future \ALERT{}s and improve overall performance.

\subsubsection{Opportunistic Mitigation on All-Bank \RFM{}}
When an \ALERT{} is raised for a row that crosses the \NBO{}, the memory controller must send an All-Bank \RFM{} (\RFMAB{}) command(s) to all DRAM banks. This is necessary because the current DRAM \ALERT{} interface cannot specify which bank issued the \ALERT{}. Consequently, the memory controller must stall all banks with the \RFMAB{} command.

An \RFMAB{} allows for opportunistic mitigations for PSQ entries in other banks, even if their activation counts are below the \NBO{}. \defense{} takes advantage of this by issuing opportunistic mitigations to the \NMIT{} highest activated rows in all banks, regardless of their activation count. This approach mitigates rows across all banks before they reach the \NBO{} and more importantly, reduces the number of future \ALERT{}s and the overall slowdown due to the mitigations.

\subsubsection{Proactive Mitigation on \REF{} commands}
Similar to TRR in DDR4, which mitigates aggressor rows during \REF{}, PRAC can benefit from proactive mitigations issued during \REF{} operations~\cite{bennett2021panopticon, UPRAC}. Proactive mitigations, like opportunistic mitigations, target the row with the highest activation count in the PSQ of each DRAM bank, regardless of whether its count exceeds \NBO{}. 
To support this, the PSQ must be at least \NMIT{} + 1 in size to handle mitigations on an \ALERT{} (\NMIT{}) and accommodate an additional entry during a \REF{} command. 
\purple{Unlike opportunistic mitigations, performing proactive mitigations for every mitigation opportunity can incur excessive energy overhead due to their higher frequency compared to \ALERT{}s. To address this, we propose an energy-optimized approach that performs mitigations \emph{only} when the activation counter of the highest activated row in the PSQ of each bank meets or exceeds the Proactive Mitigation threshold (\NPRO{}$=\frac{\text{N}_{\text{BO}}}{\text{K}}$), significantly reducing energy consumption while maintaining performance.
} 

\subsection{Sizing the Structures}
\noindent \textbf{Sizing PSQ}: A PSQ size of \NMIT{} + 1 is essential for the security of \defense{} with proactive mitigation. This size ensures that \defense{} can properly handle mitigations for an \ALERT{} (\NMIT{}) while also accommodating additional mitigation during refresh. Since the PRAC specification supports \NMIT{} values of 1, 2, and 4\cite{jedec_ddr5_prac}, we use a PSQ size of 5 for \defense{}.
\smallskip
\noindent \textbf{Sizing Counters}: Our counters are sized to avoid overflows. As per the bounds for our maximum activation count in \cref{fig:prac_trh_proa}, we set their size as $\text{minimum}(6,~\log_2(\text{\TRH{}})+1)$ bits. In practice, we use 7-bit counters for a \TRH{} of 66.
\section{Security Analysis}\label{sec:security}
We determine the Rowhammer threshold (\TRH{}) that \defense{} can securely defend against by analyzing worst-case attack patterns at different Back-Off thresholds (\NBO{}). We perform this analysis for an idealized PRAC implementation, which assumes that top-N highest activated rows are mitigated on each \ALERT{}. We then extend this to \defense{}, which uses PSQs.

\subsection{Analyzing Security of an Ideal PRAC Implementation}\label{sec:ideal_prac_security}

To model worst-case attacks on PRAC, we assume:
\begin{itemize}
    \item \textbf{Mitigation Only via \ALERT{} Back-Off:} As the JEDEC specification does not specify the policy for mitigations on \REF{}s, we bound the security of PRAC without assuming any mitigation on \REF{}s. Thus, this results in a \emph{pessimistic} upper bound on the Rowhammer threshold.
    \item \textbf{Each \ALERT{} Mitigates Top-N Activated Rows:} This models an Idealized PRAC, where each \ALERT{} mitigates the globally top-N activated rows in the bank, using `N' \RFM{}s. We show how the security guarantees for \defense{} are similar to PRAC-Ideal in \cref{sec:priority_queue_security}.
\end{itemize}

\smallskip
\subsubsection{Modeling Wave or Feinting Attack on PRAC} Similar to prior work~\cite{UPRAC}, we model the Wave or the Feinting attack~\cite{ProTRR}. This state-of-the-art attack maximizes row activations before mitigation in a PRAC-protected DRAM. This multi-round attack starts with a pool of rows, activating each row once per round. In each round, the attacker identifies and drops the mitigated rows from the pool, then uniformly activates the remaining rows. In the final round, where all remaining rows will be mitigated at the next instance, the attack focuses on hammering a single row.

The attack on PRAC consists of two phases: the \textit{Setup} phase and the \textit{Online} phase. In the Setup phase, a pool of rows of size \R{1} is generated, with each row activated \NBO{}$-1$ times to avoid mitigation. In the Online phase, the attack proceeds through multiple rounds, uniformly activating the remaining rows in each round until only a single row remains in the final round, which then receives focused hammering. Assuming the row that lasts until the final round receives a maximum of \NONLINE{} activations, the \TRH{} at which the defense is secure is:

\vspace{-0.1in}
\begin{equation} \label{eq:1}
\text{\TRH{}} > \text{\NBO{}} + \text{\NONLINE{}}
\end{equation}

\subsubsection{Bounding the Online Phase Activations} To bound the Online Phase activations (\NONLINE{}), we consider the number of attack rounds (NR), with one activation per round, followed by \ABODELAY{} + \ABOACT{} activations, which are possible before the last \ALERT{} in the final round. Additionally, the penultimate \ALERT{}'s RH mitigations can cause blast-radius (\BR{}) activations if the row in the last round is a neighbor of the mitigated rows. This increases the activations to the last row by \BR{}. 

\vspace{-0.1in}
\begin{equation} \label{eq:2}
\text{\NONLINE{}} = \text{NR} + \text{\ABOACT{}} + \text{\ABODELAY{}} + \text{\BR{}}
\end{equation}
%\vspace{-0.1in}

The number of rounds (NR) can be derived by assuming we start with a pool of \R{1} rows and recursively calculating the pool of rows (\R{N}) at each round N. In each round, the pool reduces by the number of mitigated rows (\NMIT{}) $\times$ Number-of-\ALERT{}s. Since the \RFM{}s of the last \ALERT{} in a round provide BR activations for free, each round only activates \R{N-1} - \BR{} rows, as an \ALERT{} occurs every \ABOACT{} + \ABODELAY{} activations. This allows us to determine each round's pool of rows (\R{N}).

\vspace{-0.15in}
\begin{equation} \label{eq:3}
\text{\R{N}} = \text{\R{N-1}} - \floor{\text{\NMIT{}} * (  \text{\R{N-1}} - \text{\BR{}})/ (\text{\ABOACT{}} + \text{\ABODELAY{}})}
\end{equation}

\begin{figure}[h]
\centering
\includegraphics[width=3.4in,height=\paperheight,keepaspectratio]{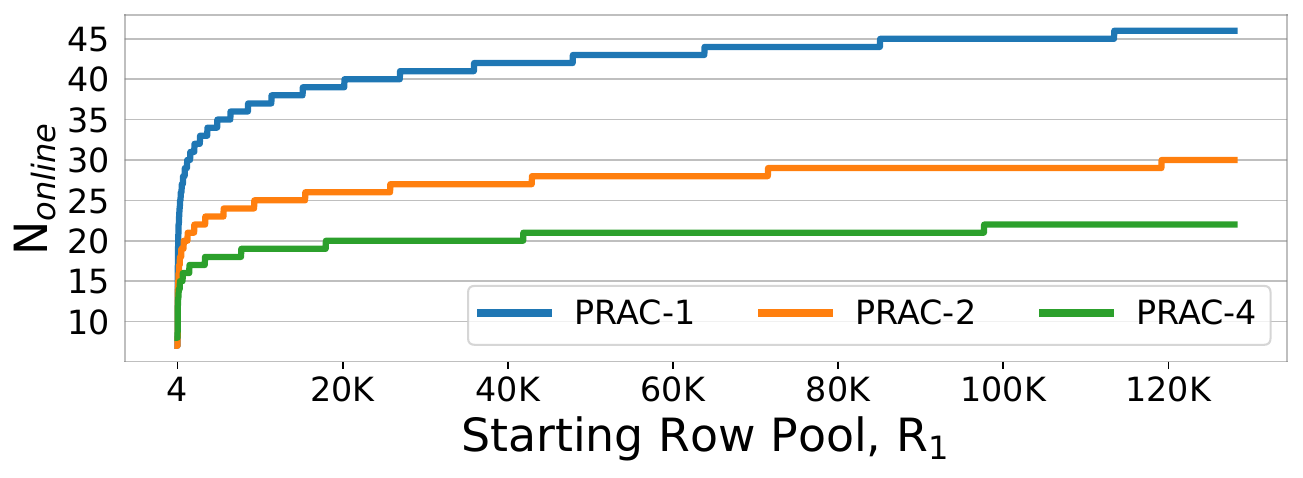}
\vspace{-0.15in}

\caption{Maximum Row Activations in Online Attack (\NONLINE{}) versus Starting Row Pool Size (\R{1}) using an analytical model. \NONLINE{} reaches a maximum of 46, 30, and 23 for PRAC-1, PRAC-2, and  PRAC-4. In empirical evaluations of the attack, our results were within 1\% of the analytical results.}
\label{fig:prac_nonline}
\vspace{-0.2in}
\end{figure}

Using \cref{eq:3}, we can recursively calculate the total number of rounds (NR), given \R{1}, and then use \cref{eq:2} to determine \NONLINE{} as \R{1} varies. As shown in \cref{fig:prac_nonline}, \NONLINE{} increases with the starting row pool size. With a maximum of 128K rows (total rows in the bank), \NONLINE{} can reach 46 for PRAC-1 (\NMIT{} = 1), 30 for PRAC-2 (\NMIT{} = 2), and 23 for PRAC-4 (\NMIT{} = 4).

\subsubsection{Constraint on \R{1} Due to Attack Time}
\begin{figure}[ht]
\vspace{-0.1in}
\centering
\includegraphics[width=3.4in,height=\paperheight,keepaspectratio]{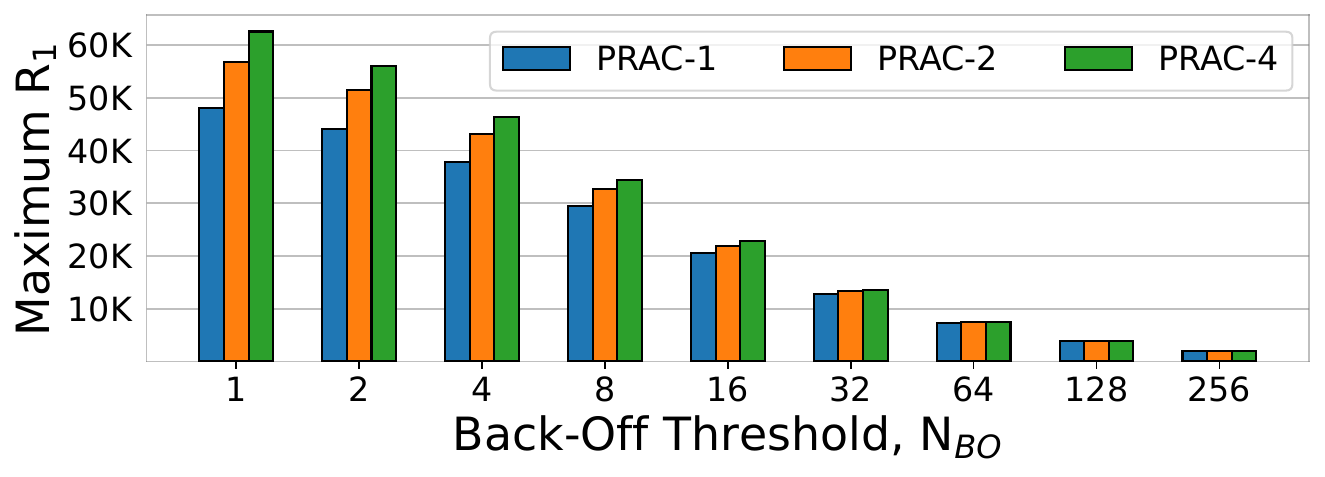}
\vspace{-0.1in}
\caption{Maximum Starting Row Pool (\R{1}) versus Back-Off Threshold (\NBO{}). As \NBO{} increases, the maximum possible \R{1} decreases. This is because the time taken by the setup phase increases at higher \NBO{}.}
\label{fig:prac_setup}
\vspace{-0.05in}
\end{figure}

The starting pool size of the attack (\R{1}) is constrained by the time required for both the Setup and Online phases. The Setup phase involves activating \R{1} rows \NBO{}-1 times each, and both phases must be completed within \TREFW{} (32ms) for the attack to succeed, limiting \R{1}. 
\cref{fig:prac_setup} shows the maximum \R{1} as \NBO{} varies. 
At \NBO{} of 1, the Setup phase requires negligible time, and \R{1} is limited by the Online phase duration, ranging from 50K to 62K for PRAC-1 to PRAC-4. 
As \NBO{} increases to 256, the maximum \R{1} size drops to 2K due to the Setup phase dominating. As we move from PRAC-1 to PRAC-4, the time required for the Online phase decreases for a given \R{1} since the number of mitigations per \ALERT{} increases. Consequently, for the same \NBO{}, \R{1} increases from PRAC-1 to PRAC-4.

\subsubsection{Quantifying \TRH{}} 
\begin{figure}[ht]
\centering
\vspace{-0.1in}
\includegraphics[width=3.4in,height=\paperheight,keepaspectratio]{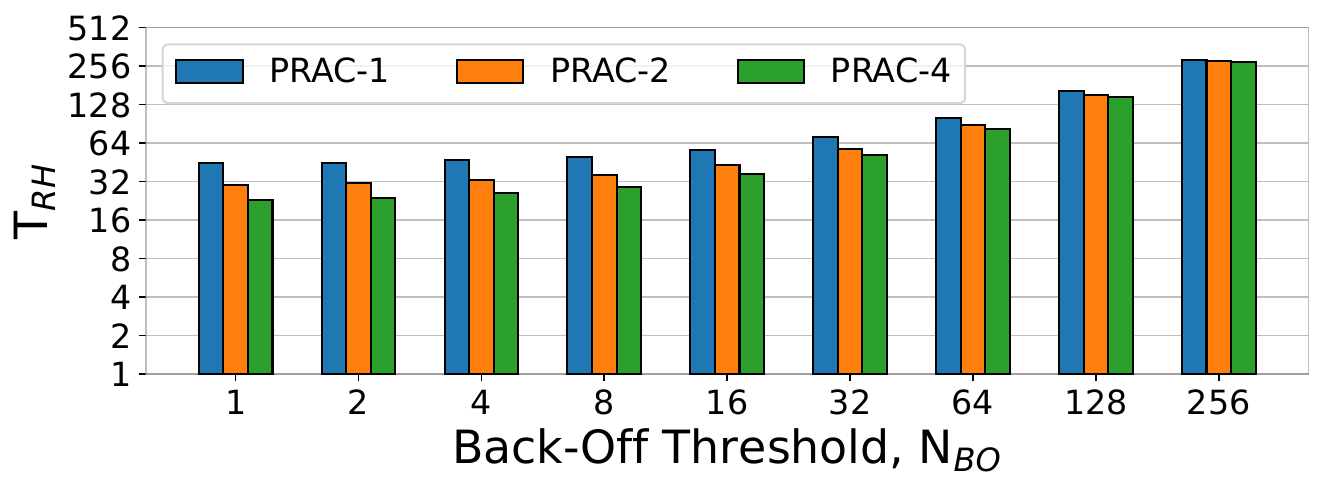}
\vspace{-0.1in}
\caption{\TRH{} values for which PRAC-N is secure as \NBO{} varies. At \NBO{} of 1, the lowest possible \TRH{} for PRAC-1, 2, and 4 is 44, 29, and 22, respectively.}
\label{fig:prac_trh}
\vspace{-0.1in}
\end{figure}

Using the constraints on \R{1} for different \NBO{} from \cref{fig:prac_setup} and \cref{fig:prac_nonline}, we can determine the maximum \NONLINE{} value and subsequently the \TRH{} that PRAC can tolerate, using \cref{eq:1}. \cref{fig:prac_trh} shows the lowest possible \TRH{} for which PRAC is secure at different \NBO{}s. At \NBO{} of 1, PRAC-1, PRAC-2, and PRAC-4 are secure for \TRH{} values of 44, 29, and 22, respectively. As \NBO{} increases, the value of \NONLINE{} remains relatively unchanged. At \NBO{} of 256, the securely mitigated \TRH{} values are 289, 279, and 274 for PRAC-1, PRAC-2, and PRAC-4, respectively.

A similar analysis in prior work, UPRAC~\cite{UPRAC}, fails to account for the activations in the last round for a single row and the effect of transitive attack mitigation. This can exacerbate the attack and the Setup phase time, which bounds the pool of rows in the attack. Unlike prior claims~\cite{UPRAC} that PRAC-1 to PRAC-4 are secure at a minimum \TRH{} of 17 to 10, our precise modeling shows that they are secure only up to \TRH{} of 44 to 22, respectively.

\subsection{Effect of Priority-Based Service Queue in \defense{}}\label{sec:priority_queue_security}
Unlike an `Ideal' PRAC, \defense{} uses a priority-based service queue (PSQ) of $N$ entries ($N \geq \text{\NMIT{}}$) to determine the top \NMIT{} rows to mitigate when an \ALERT{} is raised. 
So, we seek to answer two questions: 
(1) Is the PSQ secure from attacks that exploit full FIFO queues?
(2) Is the PSQ vulnerable to any new attack patterns due to limited queue capacity?

\begin{figure}[htb]
\vspace{-0.1in}
\centering
\includegraphics[width=3.4in,height=\paperheight,keepaspectratio]
{"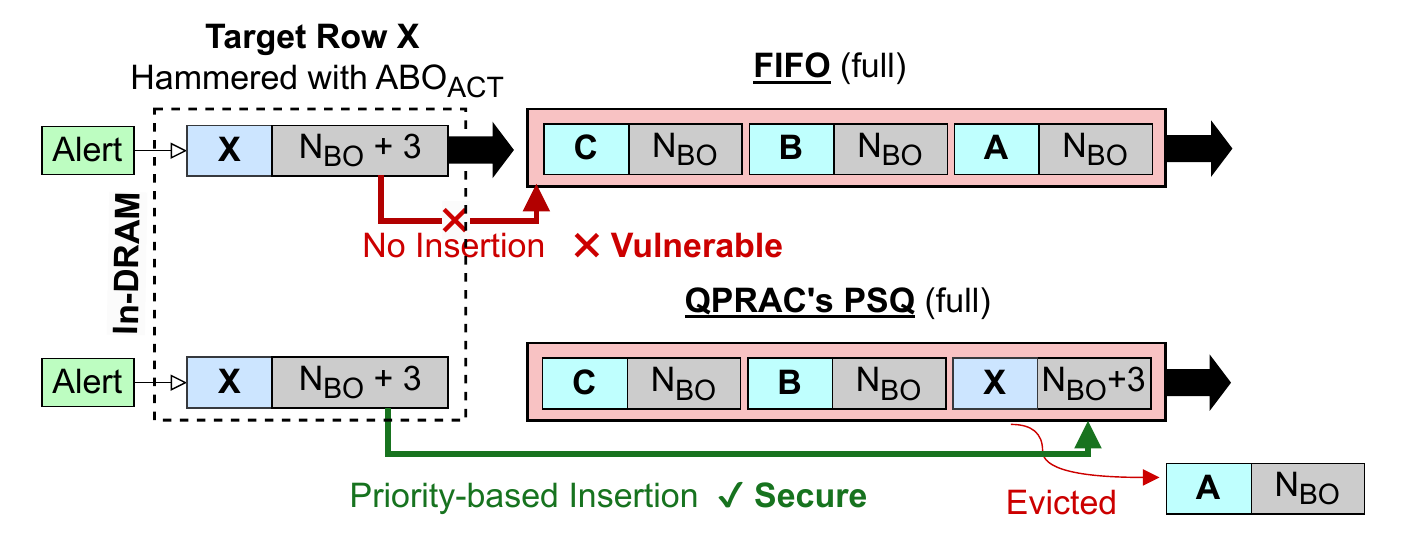"}
\vspace{-0.05in}
\caption{FIFO-based queues vs \defense{}'s PSQ. FIFO is vulnerable to insertion bypass when full, while PSQ uses priority-based insertion, prioritizing rows with higher activation counts to secure against \ABOACT{}-based hammering.}
\label{fig:fifovsprac}
\end{figure}

\noindent\textbf{Tolerating Full PSQ}: The PSQ design addresses the security limitations of FIFO-based queues when they are full. While FIFO is vulnerable to attacks (\cref{sec:attack2}-\Attacktwo{}), where highly activated rows hammered with \ABOACT{} are bypassed when the queue is full, as shown in \cref{fig:fifovsprac}, priority-based insertions in QPRAC insert such rows even when PSQ is full. This effectively tracks and mitigates such rows, making \defense{} secure from such attacks.

\noindent\textbf{Security Holds Under Size Constraint}: A size-constrained PSQ, however, means it cannot always hold the globally top $N$ rows. Consequently, mitigations can target rows with the highest global activation counts or local maxima.
Yet, under the wave attack discussed in \cref{sec:ideal_prac_security}, all mitigation decisions executed through the PSQ consistently target the globally most frequently activated rows, thus aligning with the `Ideal' PRAC under this attack.

\begin{figure}[htb]
\centering
\includegraphics[width=3.6in,height=\paperheight,keepaspectratio]
{"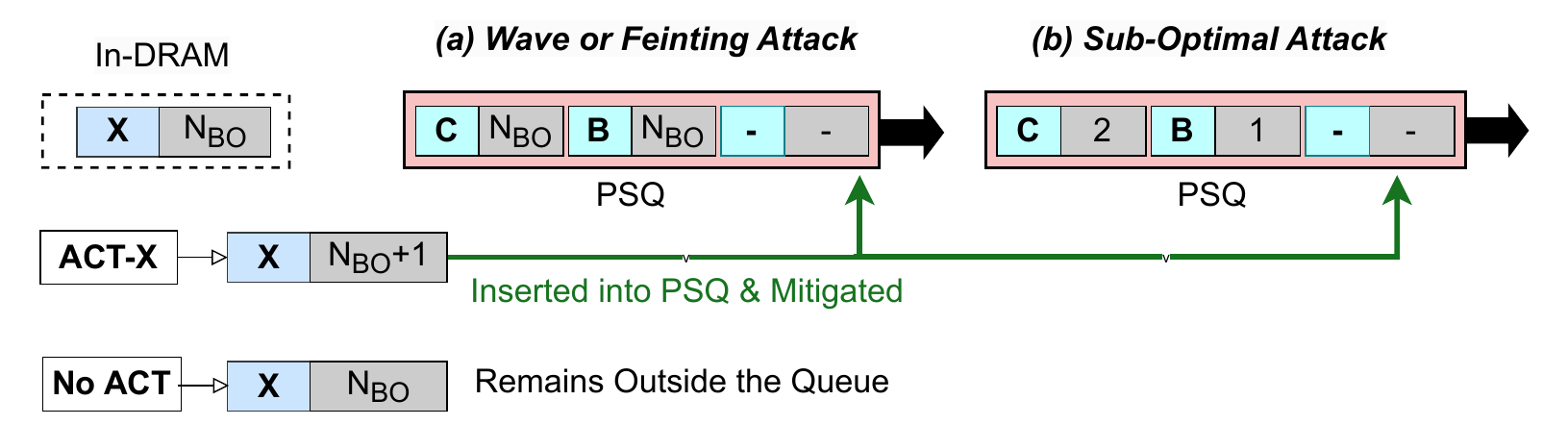"}
\vspace{-0.15in}
\caption{\defense{}'s tracking of global maximums. Sometimes, the global maximum may be outside the PSQ, but it cannot be activated further without being inserted into the queue and mitigated.}
\label{fig:topk}
\end{figure}

This is because the wave attack uniformly activates pool rows to the same maximum activation count. Even if some rows are evicted from the PSQ due to insufficient capacity, as demonstrated in \cref{fig:topk}(a), they are reinserted into the PSQ the next time they are activated. Therefore, during attacks, including but not limited to the wave attack, the PSQ predominantly holds the top-$N$ most frequently activated rows. Simulations of the Wave or Feinting attack show that the maximum activation counts for \defense{} (with PSQ) are identical to those of the ideal PRAC (without PSQ).

\textbf{Alternative Attacks Are Inferior}: Any attack variant that forces mitigations on locally (not globally) maximum activation counts is sub-optimal. This happens only if (1) the global maximum is evicted from the PSQ and not activated again, and (2) other rows with fewer activation counts are activated, inserted into the PSQ, and mitigated on \ALERT{}. Such an attack is sub-optimal because, as shown in \cref{fig:topk}(b), the globally maximum activated row cannot increase its activation outside the PSQ. Time spent keeping the global maximum row outside of the PSQ is wasted. When activated again, the global maximum row is reinserted into the PSQ and preserved until mitigation. Thus, the PSQ \emph{does not} introduce worse attacks.

\subsection{Effect of Proactive Mitigation on \defense{}}\label{sec:proactive_security}
\defense{} can be co-designed with proactive mitigation on \REF{}s, where one mitigation per \REF{} is proactively issued for the highest activated row in the PSQ, as long as the queue size is at least \NMIT{}$ + 1$. This approach has two key benefits: (1) reducing the number of rows that can reach \NBO{} in the \textit{Setup} phase, and (2) mitigating more rows per round in the \textit{Online} phase, thereby reducing the number of attack rounds.

\subsubsection{Impact on Setup Phase} 
\begin{figure}[htb]
\vspace{-0.1in}
\centering
\includegraphics[width=3.4in,height=\paperheight,keepaspectratio]{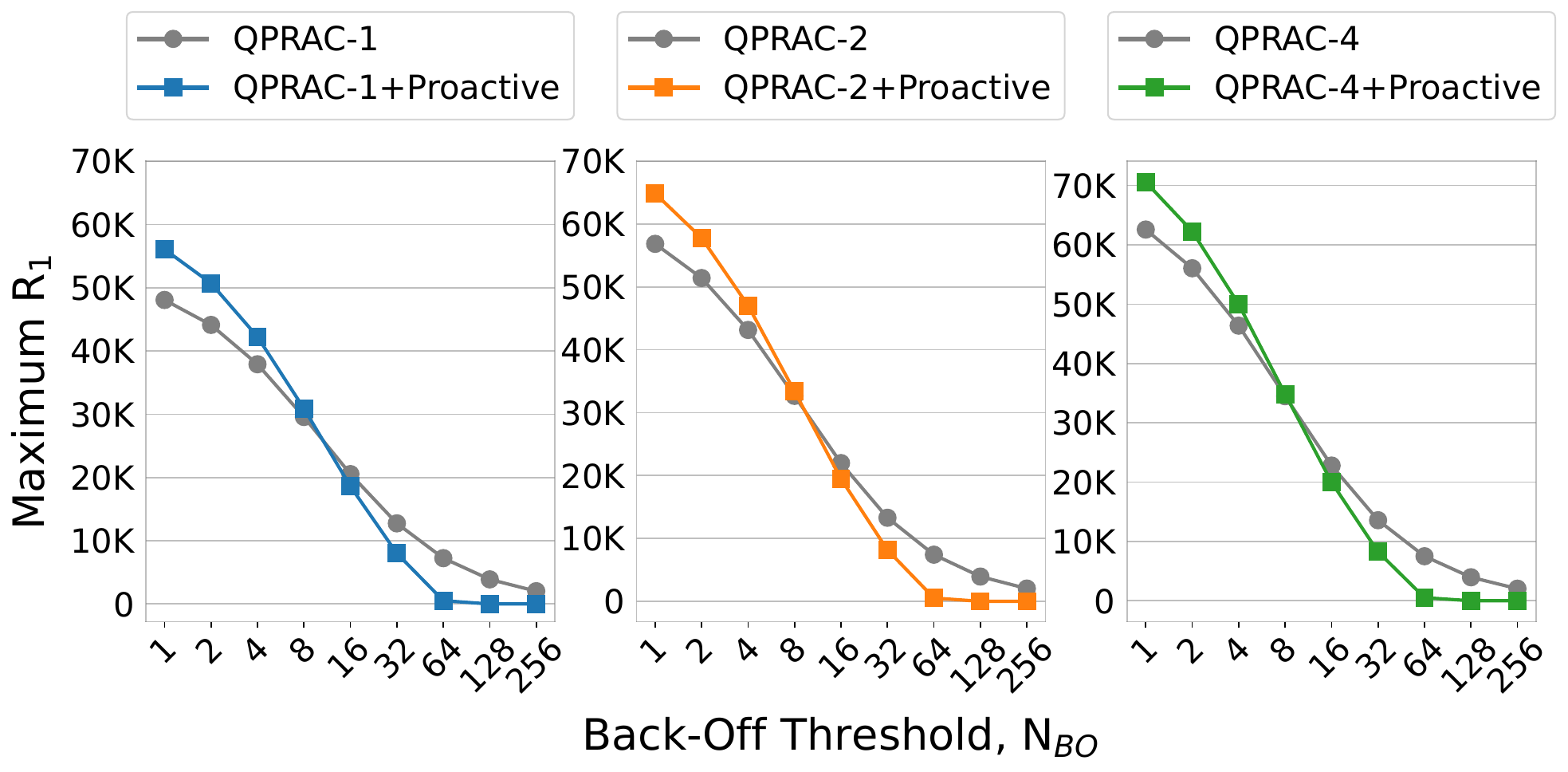}
\vspace{-0.1in}
\caption{Maximum Starting Row Pool Size (\R{1}) in the wave attack for \defense{} with proactive mitigation compared to QPRAC without proactive mitigation (labeled simply as \defense{}). For higher \NBO{}, where the Setup phase consumes more time, the \R{1} size reduces considerably due to proactive mitigation.}
\label{fig:prac_setup_proa}
\vspace{-0.1in}
\end{figure}

Rows are uniformly activated to reach \NBO{} - 1 activations during the Setup phase. Proactive mitigations reduce the pool of rows (\R{1}) available for the attack. The number of mitigations ($M$) is calculated as the number of activations ($A$) in the Setup phase divided by the number of activations per \TREFI{} (67), as $M = \frac{A}{67}$. As shown in \cref{fig:prac_setup_proa}, with proactive mitigation, the pool of rows (\R{1}) available for the attack decreases. For \NBO{} values of 16 and higher, proactive mitigation significantly reduces \R{1} compared to PRAC alone, with \NBO{} values of 128 and 256 completely defeating the attack by mitigating the entire \R{1} pool before it crosses \NBO{}. For \NBO{} lower than 16, the short Setup phase does not experience as many mitigations; the online phase being shorter with the proactive mitigations allows for larger values of \R{1} to be possible for the attack.

\subsubsection{Impact on Online Phase} 
\begin{figure}[htb]
\centering
\includegraphics[width=3.4in,height=\paperheight,keepaspectratio]{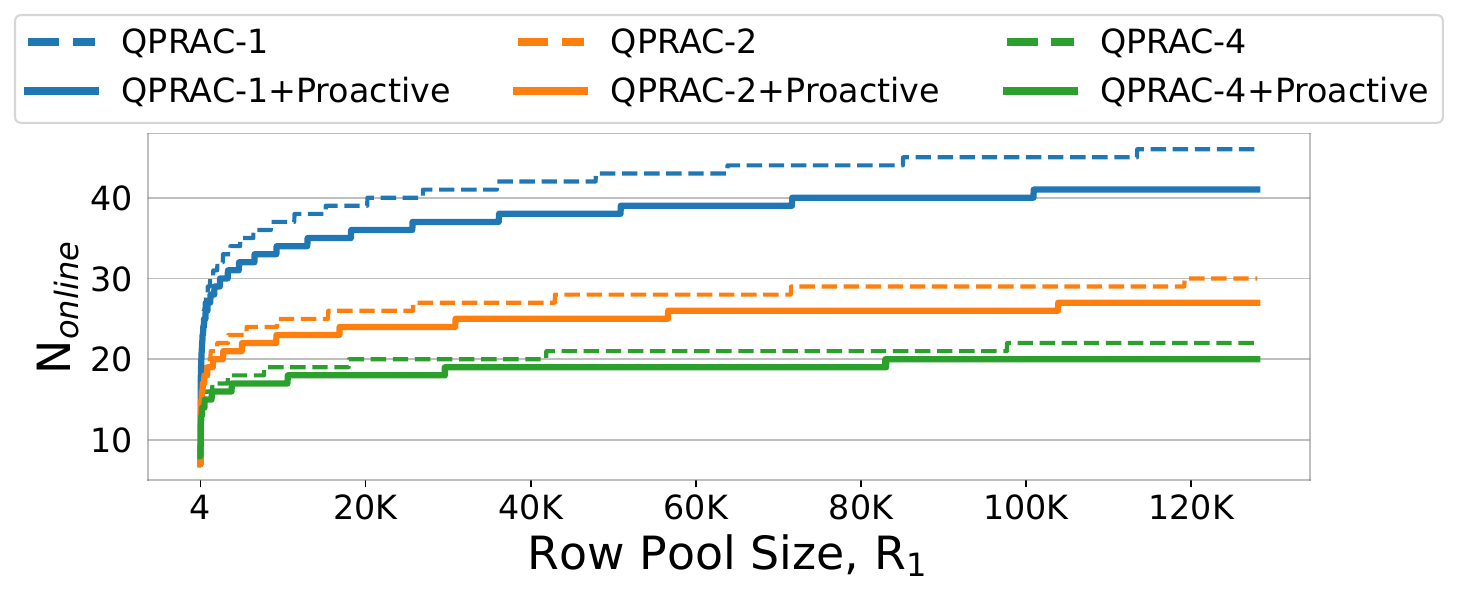}
\vspace{-0.1in}
\caption{Maximum Activations Per Row in Online Phase (\NONLINE{}) for \defense{} \emph{with} proactive mitigation versus \defense{} \emph{without} proactive mitigation (labeled simply as \defense{}). \NONLINE{} decreases by a maximum of 5, 2, and 1 for \defense{}-1, \defense{}-2, and \defense{}-4 with proactive mitigations, respectively.}
\label{fig:prac_nonline_proa}
\vspace{-0.2in}
\end{figure}

In the Online Phase, the pool of rows in each round decreases more rapidly due to proactive mitigations. The number of additional rows mitigated is calculated by dividing the total Onlie phase time (Activation-Time + \ALERT{}-Time) by \TREFI{}. Including these in \cref{eq:3}, we calculate \NONLINE{} versus $R_1$, shown in \cref{fig:prac_nonline_proa}.

\begin{figure}[h]
\vspace{-0.1in}
\centering
\includegraphics[width=3.4in,height=\paperheight,keepaspectratio]{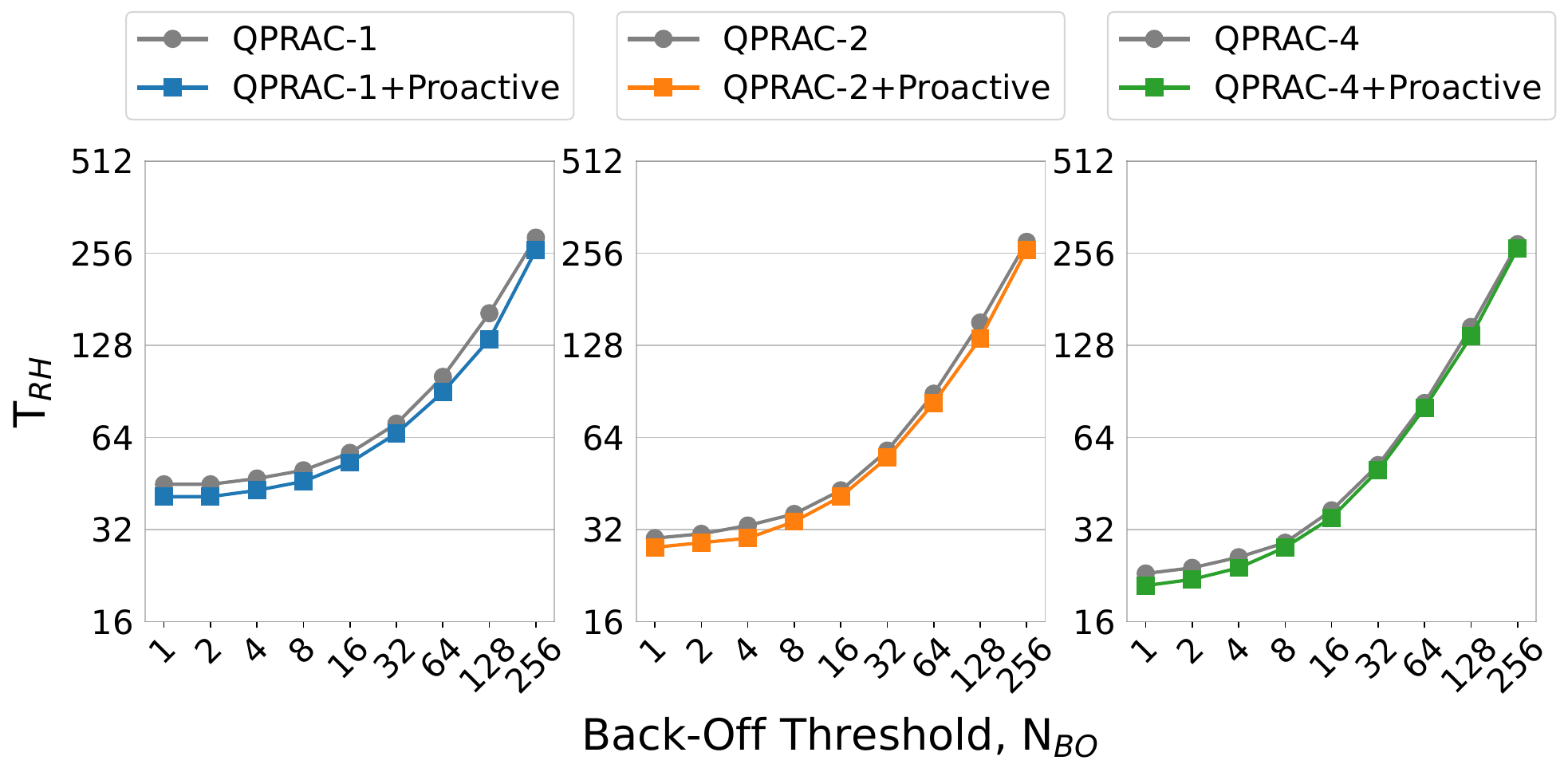}
\vspace{-0.1in}
\caption{The \TRH{} values for QPRAC with proactive mitigation and without proactive mitigation (labeled simply as QPRAC). With proactive mitigation, the lowest possible \TRH{} at \NBO of 1 is 40, 27, and 20 for QPRAC-1, 2, and 4, respectively. In contrast, the lowest possible \TRH{} without proactive mitigation is 44, 29, and 22 for QPRAC-1, 2, and 4, respectively.}
\label{fig:prac_trh_proa}
\vspace{-0.1in}
\end{figure}

Using \R{1} from \cref{fig:prac_setup_proa} and the associated \NONLINE{} from \cref{fig:prac_nonline_proa}, we can determine the minimum \TRH{} supported by \defense{} with proactive mitigations, as shown in \cref{fig:prac_trh_proa}. For \NBO{} of 1, the minimum supported \TRH{} drops to 40, 27, and 20 for PRAC-1, 2, and 4, respectively, with proactive mitigation, compared to 44, 29, and 22 without proactive mitigation. For our default \NBO{} of 32, proactive mitigation can defend against a \TRH{} of 66, 55, and 50 for \defense{}-1, 2, and 4, respectively, compared to 71, 58, and 52 without proactive mitigation.

Our energy-aware design, \defense{} with 
energy-aware proactive mitigations (\defense{}+Proactive-EA), which skips some wasteful proactive mitigations, achieves a security level between \defense{} and \defense{}+Proactive. This is because it also reduces the number of rows available in the \textit{setup} phase but to a lesser extent than proactive mitigation on every \REF{}.

\section{Evaluation Methodology}\label{eval}
\begin{figure*}[htb]
    \centering
    \includegraphics[width=0.9\linewidth]{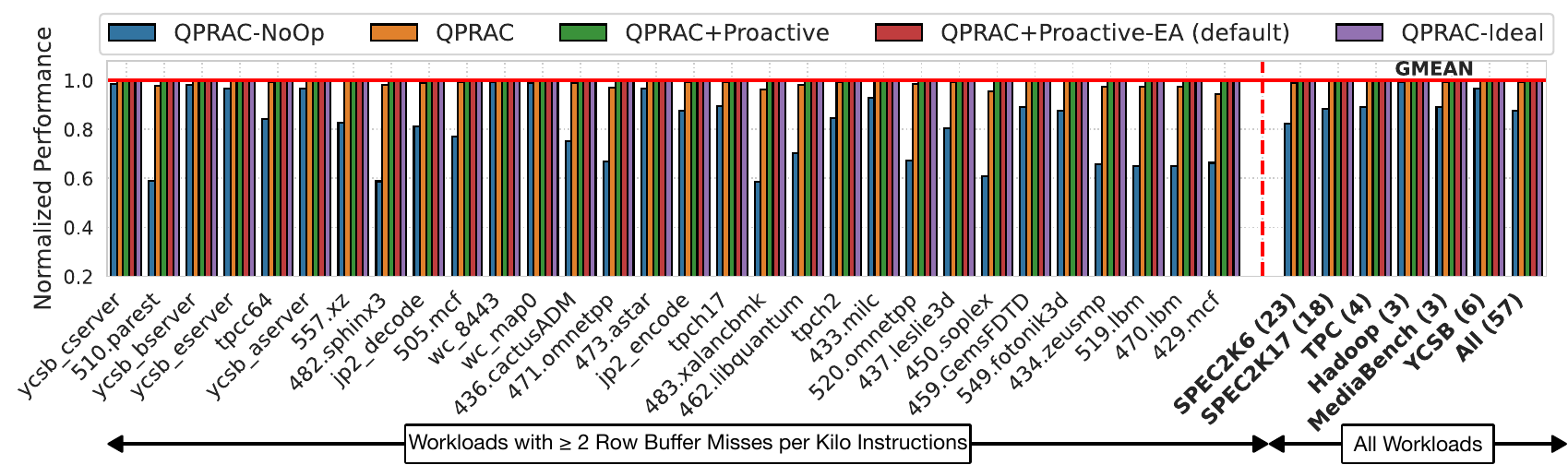}
    \vspace{-0.1in}
    \caption{Normalized performance of QPRAC with a 5-entry priority queue at a Back-Off threshold (\NBO{}) of 32 and 1 \RFM{} per \ALERT{} Back-Off (ABO), compared to an insecure baseline without ABO. \defense{}-NoOp incurs a considerable 12.4\% slowdown on average. In contrast, other \defense{} implementations with opportunistic mitigations--\defense{}, \defense{}+Proactive, \defense{}+Proactive-EA (our default), and \defense{}-Ideal--result in negligible 0.8\% slowdown for \defense{}, and no slowdown for the latter three due to proactive mitigations.
    }
    \label{fig:result:perf}
\end{figure*}
\begin{figure*}[t]
    \vspace{-0.15in}
    \centering
    \includegraphics[width=0.9\linewidth]{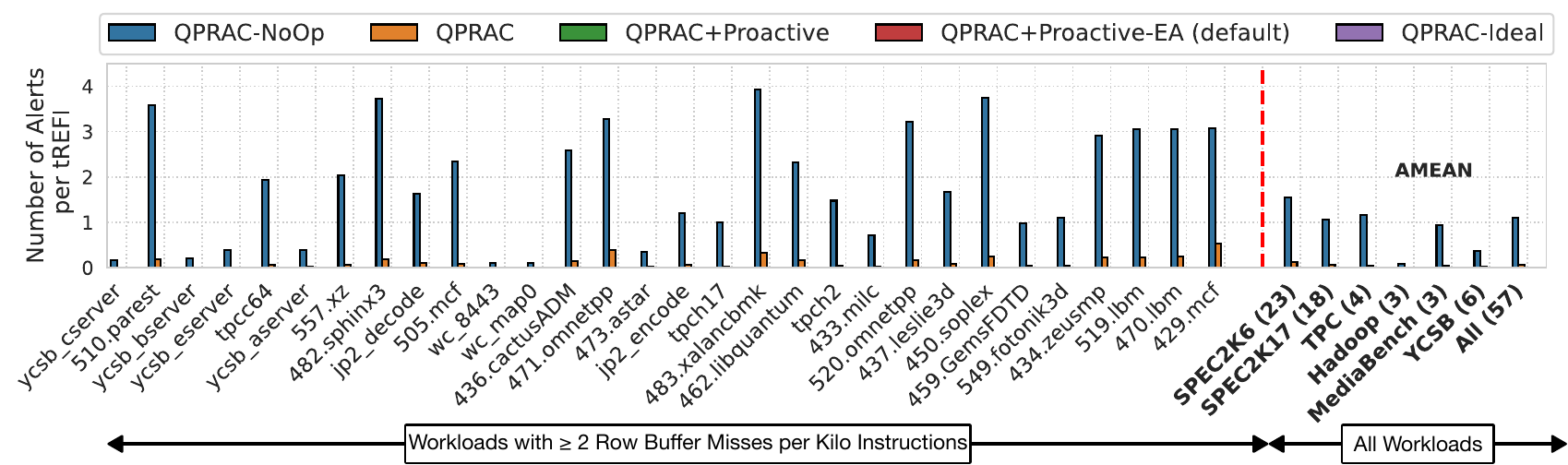}
    \vspace{-0.1in}
    \caption{The frequency of \ALERT{} Back-Off (ABO) occurrences per \TREFI{} interval for different \defense{} implementations at a Back-Off threshold of 32 and 1 \RFM{} per ABO. \defense{}-NoOp experiences nearly 1.1 ABO per \TREFI{}. In contrast, \defense{} (with opportunistic mitigation), \defense{}+Proactive, \defense{}+Proactive-EA (our default), and \defense{}-Ideal have insignificant 0.07 ABO per \TREFI{}, with no ABO occurrences for the latter three due to proactive mitigations.
    }
    \label{fig:result:num_alert}
    \vspace{-0.15in}
\end{figure*}

\noindent \textbf{Simulation Framework}: We evaluate designs using the cycle-accurate trace-based DRAM simulator Ramulator2~\cite{kim2015ramulator,ramulator2}. We use an out-of-order core model in Ramulator2, similar to prior RH works~\cite{yauglikcci2021blockhammer, comet, olgun2023abacus, HIRA, rowpress, UPRAC, dapper}. Our system configuration is shown in Table~\ref{table:system_config}. We simulate a baseline system with a 4-core, 8MB shared LLC equipped with 64GB DDR5 memory (one channel, two ranks). The memory is configured using timing parameters based on the Micron 32Gb DDR5 device~\cite{micron_ddr5}, including PRAC-specific timing changes~\cite{jedec_ddr5_prac}. Each DRAM bank consists of 128K rows, each 8KB in size. Within a 32ms refresh window, a single bank can undergo up to approximately 550K activations.

\begin{table}[h!]
\vspace{-0.15in}
\begin{center}
\begin{small}
\caption{System Configuration}{
\resizebox{0.95\columnwidth}{!}{
\begin{tabular}{|c|c|}
\hline
  Out-Of-Order Cores           &  4 Core, 4GHz, 4 wide, 352 entry ROB          \\
  Last Level Cache (Shared)    & 8MB, 8-Way, 64B lines \\ \hline
  Memory Size, Type                  & 64 GB,  DDR5 \\
  Bus Speed             & 3200MHz (6400MHz DDR) \\
  DRAM Organization      & 4 Bank x 8 Groups x 2 Ranks x 1 Channel \\
  tRCD, tCL, tRAS & 16ns, 16ns, 16ns\\
  tRP, tRTP, tWR, \TRC{} & 36ns, 5ns, 10ns, 52ns \\ 
  \TRFC{}, \TREFI{}     &   410 ns, 3.9$\mu$s \\
  t\ABOACT{}, t\RFMAB{} & 180ns, 350ns \\
  Rows Per Bank, Size                 & 128K, 8KB \\ \hline
\end{tabular}}
\vspace{-0.1in}
\label{table:system_config}
}
\end{small}
\end{center}
\vspace{-0.15in}
\end{table}

\noindent \textbf{Evaluated Designs}: We compare \defense{} against a baseline DRAM that also uses DDR5 PRAC timings but without the \ALERT{} Back-Off (ABO) based mitigations. We extend Ramulator2 to faithfully model the per-row activation counters, the ABO protocol, and their timing constraints. We evaluate the following \defense{} configurations: 1) \textbf{\defense{}-NoOp} that performs mitigation on an \RFM{} \emph{only} for the bank with the entry that reached the Back-Off Threshold (\NBO{}). 2) \textbf{\defense{}} that mitigates the highest activated row(s) in the priority-based service queue (PSQ) from every bank when an \RFM{} is received. 3) \textbf{\defense{}+Proactive} that additionally performs proactive mitigation for the highest activated entry during the refresh operations (\REF{}) for each bank. 
\purple{4) \textbf{\defense{}+Proactive-EA} is an energy-aware extension of \defense{}+Proactive. It mitigates the most frequently activated rows in the PSQ during proactive mitigations only when their counter reaches \NPRO{}. By default, we set \NPRO{} to half of \NBO{}.
}
5) \textbf{\defense{}-Ideal} is an ideal implementation that knows and mitigates the `top-N' highest activated rows for each ABO in addition to proactive mitigations, similar to UPRAC~\cite{UPRAC}.

\noindent \textbf{Workloads}: We use 57 applications from SPEC2006~\cite{SPEC2006}, SPEC2017~\cite{SPEC2017}, TPC~\cite{TPC}, Hadoop~\cite{hadoop}, MediaBench~\cite{MediaBench}, and YCSB~\cite{ycsb} benchmarks, open-sourced with Ramulator2~\cite{ramulator_opensource}. We run four homogeneous workloads until each core completes 500 million instructions.
We use a default Back-Off Threshold (\NBO{}) of 32 and 1 \RFM{} per \ALERT{} (PRAC-1). We also vary \NBO{} (128 to 16), the number of \RFM{}s per \ALERT{} (1, 2, or 4), and queue sizes (1 to 5). We use the weighted speedup to evaluate the performance of \defense{} designs. 

\section{Results and Analysis}

\subsection{Performance Overhead}
Figure~\ref{fig:result:perf} shows the performance overhead of \defense{} implementations normalized to a non-secure baseline DDR5 system. \defense{}-NoOp incurs a 12.4\% slowdown, while \defense{}, \defense{}+Proactive, \defense{}+Proactive-EA (our default), and \defense{}-Ideal experience minimal overheads of 0.8\%, and no overhead for the latter three with proactive mitigations.

The higher overhead for \defense{}-NoOp arises because it only mitigates DRAM banks that reach the Back-Off threshold (\NBO{}) when these banks receive the \RFM{}s caused by Alert Back-Off (ABO). Consequently, when other banks reach the \NBO{}, they issue separate \ALERT{}s followed by \RFM{}s, reducing effective DRAM bandwidth. Opportunistically mitigating all PSQs on each \RFM{}, as implemented in \defense{} by default, significantly reduces the number of \ALERT{}s. \defense{}+Proactive further improves performance by leveraging proactive mitigations to reduce the burden on \ALERT{}s, bringing no additional performance overhead. QPRAC+Proactive-EA maintains optimal performance as its \textit{proactive mitigation threshold} (\NPRO) still ensures aggressor rows are mitigated well before they reach \NBO{}, thus ensuring minimal \ALERT{}s and avoiding their slowdown. \defense{}-Ideal (which similarly has proactive mitigation) shows identical performance to \defense{}+Proactive-EA, underscoring the effectiveness of our design.

Figure~\ref{fig:result:num_alert} shows the \ALERT{}s per \TREFI{} for our \defense{} implementations. \defense{}-NoOp incurs almost one \ALERT{} every \TREFI{}, leading to considerable performance degradation. It drops more than 20\% performance in many memory-intensive applications, such as \emph{429.mcf} and \emph{482.sphinx3}, and a maximum of 46\% performance drops in \emph{510.parest} due to frequent \ALERT{}s, occurring more than two Alerts per \TREFI{}. In contrast, \defense{}, which has opportunistic mitigations, significantly reduces the number of \ALERT{}s to 0.07 per \TREFI{}. Finally, \defense{}+Proactive-EA, our default design, incurs no overhead because proactive mitigation during \REF{} eliminates the \ALERT{}s.

\subsection{Sensitivity to Number of \RFM{}s per \ALERT{}}
\begin{figure}[b!]
    \centering    \includegraphics[width=0.4\textwidth,height=\paperheight,keepaspectratio]{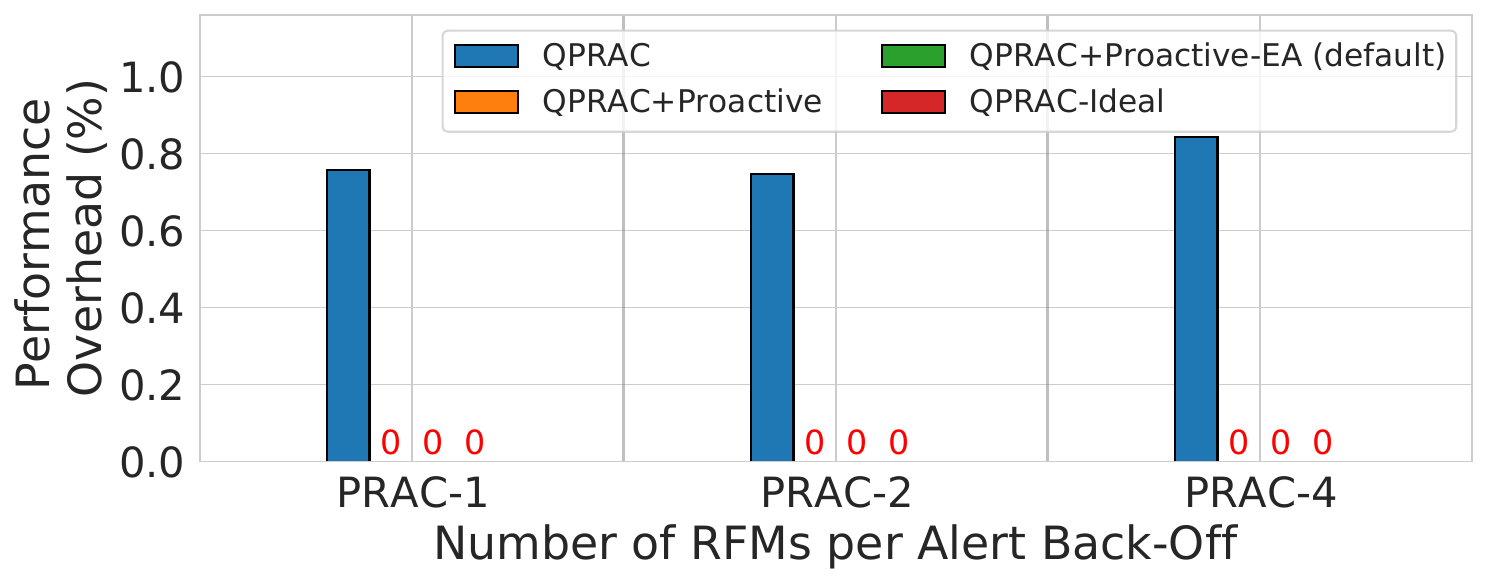}
    \vspace{-0.1in}
    \caption{Slowdown of \defense{} for \RFM{}s/\ALERT{} values of 1, 2, and 4 (default = 1). \defense{} experiences a slowdown of 0.8\% to 0.9\%, while 
 \defense{} with proactive mitigations--\defense{}+Proactive, \defense{}+Proactive-EA, and \defense{}-Ideal--incur no overhead for 1 to 4 \RFM{}s per \ALERT{}.}
    \label{fig:result:vayring_num_RFMs}
\end{figure}

PRAC specification allows a predefined number of All-Bank \RFM{}s (1, 2, or 4) to be issued per \ALERT{}. Our default is 1 \RFM{} per \ALERT{}, but using 2 or 4 \RFM{}s per \ALERT{} can further reduce \TRH. Figure~\ref{fig:result:vayring_num_RFMs} shows the performance overheads of \defense{} under PRAC-1, PRAC-2, or PRAC-4 configurations (1, 2, or 4 \RFM{}s per \ALERT{}) compared to the baseline. \defense{} alone incurs 0.8\%, 0.8\%, and 0.9\% slowdown at 1, 2, and 4 \RFM{}s per \ALERT{} respectively. In contrast, QPRAC with proactive mitigations--\defense{}+Proactive, \defense{}+Proactive-EA, and \defense{}-Ideal--incur no overhead, as proactive mitigations during REFs eliminate ABO occurrences.

The performance overhead remains similar across PRAC-1 to PRAC-4. This is because, although \RFM{}s per \ALERT{} from 1 to 4 increases the mitigation cost per \ALERT{} (\textit{i.e.}, the banks are blocked for a longer period), it also reduces the \ALERT{} frequency significantly. For example, compared to PRAC-1, PRAC-2 and PRAC-4 decrease \ALERT{}s by 1.9$\times$ and 3.3$\times$, respectively. As a result, the slowdowns remain consistent.
  
\subsection{Sensitivity to Service Queue Size}
\begin{figure}[ht!]
\vspace{-0.1in}
    \centering    \includegraphics[width=0.45\textwidth,height=\paperheight,keepaspectratio]{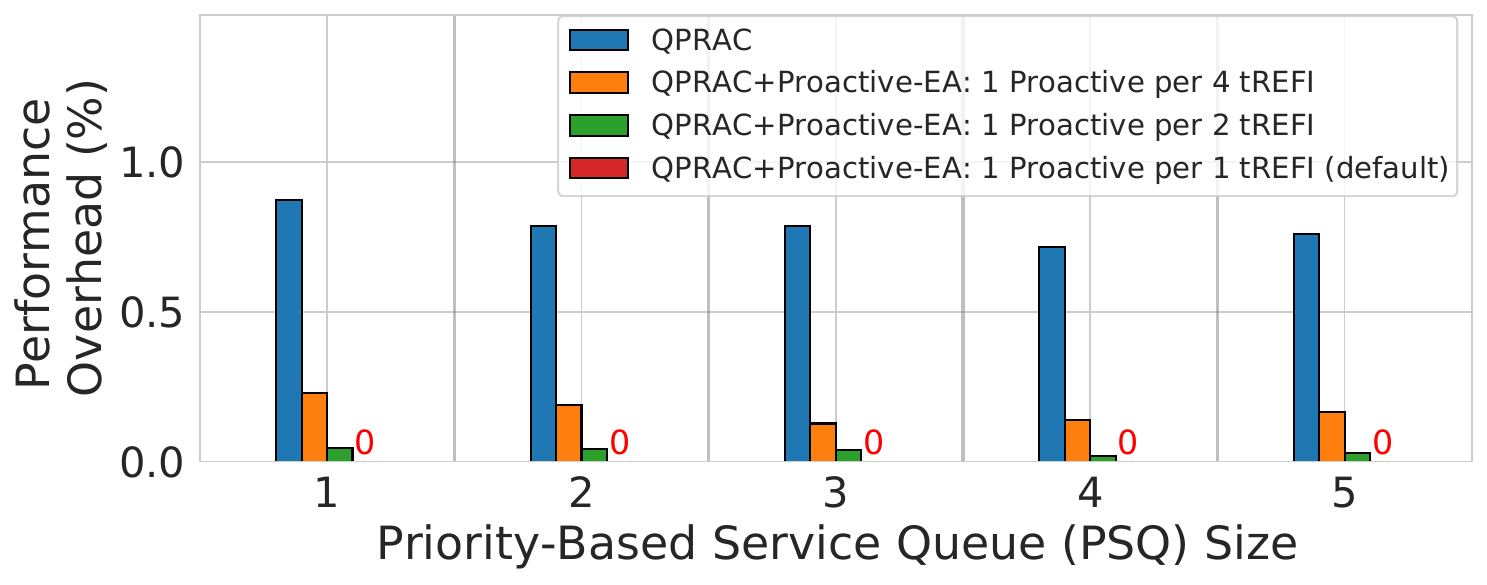}
    \vspace{-0.1in}
    \caption{Slowdown of \defense{} as the queue size varies for different proactive mitigation frequencies. \defense{} shows negligible overhead ($<$ 1\%) across all evaluated queue sizes, with slightly better performance at larger sizes.
    }
    \label{fig:eval_queue_sizes}
    \vspace{-0.15in}
\end{figure}

\cref{fig:eval_queue_sizes} shows the \defense{} performance as the priority-based service queue (PSQ) sizes vary from 1 to 5. \defense{}'s performance remains consistently low, with less than 1\% overhead across all evaluated queue sizes, showing slightly better performance at larger sizes. A 5-entry queue is used by default for \defense{} to ensure compatibility with the PRAC specification, supporting up to a \NMIT{} of 4 (i.e., PRAC-4).

\subsection{Sensitivity to Back-Off Threshold}
\begin{figure}[h!]
    \vspace{-0.1in}
    \centering
    \includegraphics[width=0.45\textwidth,height=\paperheight,keepaspectratio]{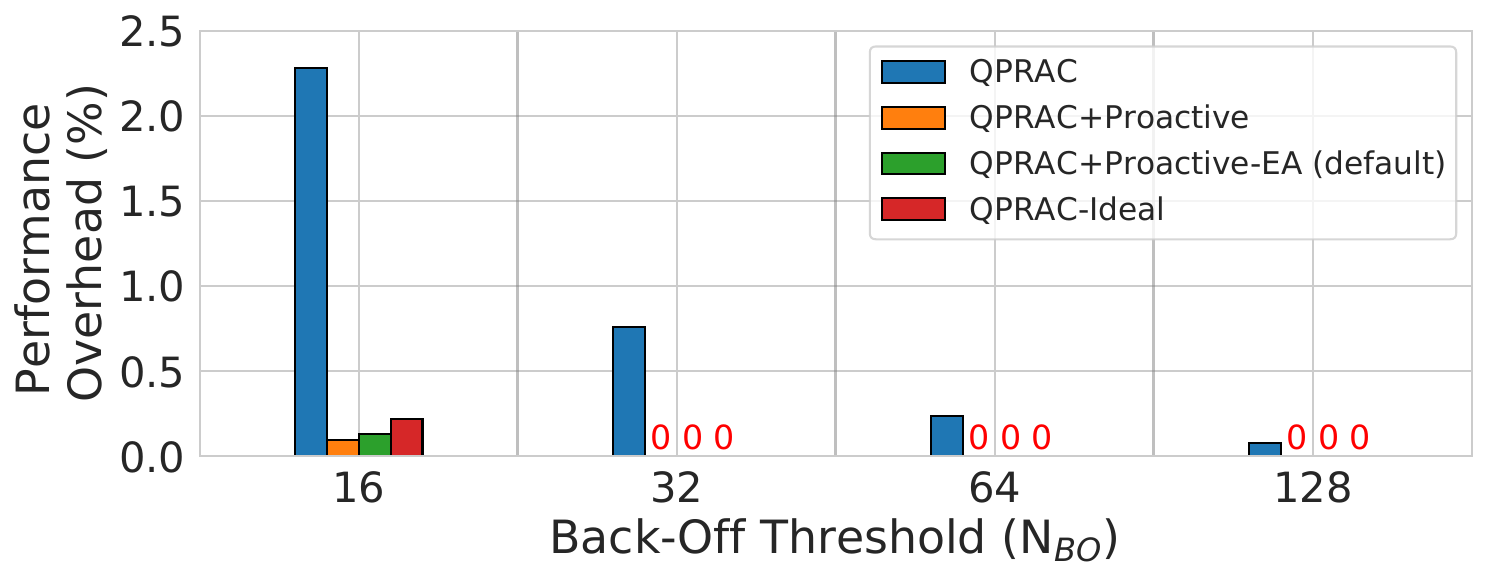}
    \vspace{-0.1in}
    \caption{Performance overhead of \defense{} as the Back-Off Threshold (\NBO{}) varies. \defense{} incurs a 0.8\% slowdown at \NBO{} of 32 and 2.3\% at \NBO{} of 16. In contrast, \defense{} designs with proactive mitigations show negligible slowdown across all evaluated \NBO{} values, with no overhead at \NBO{} of 32 and less than 0.3\% at \NBO{} of 16.
    }
    \vspace{-0.15in}
    \label{fig:result:sensNBO}
\end{figure}

\cref{fig:result:sensNBO} shows the performance of \defense{} as the Back-Off threshold (\NBO{}) varies from 16 to 128. Higher \NBO{} values reduce the number of \ALERT{}s and limit slowdown but also increase the \TRH{} tolerated by the defense. For \NBO{} of 32 or more, \defense{} incurs negligible slowdown of less than 0.8\%, while our designs with proactive mitigations, \defense{}+Proactive and \defense{}+Proactive-EA (our default), incur no slowdown, similar to \defense{}-Ideal. 
At \NBO{} of 16, \defense{} has 2.3\% slowdown, while \defense{}+Proactive, \defense{}+Proactive-EA, and \defense{}-Ideal, have less than 0.3\% slowdowns. Although reducing \NBO{} from 32 to 16 increases slowdowns for \defense{} from 0.8\% to 2.3\%, it only reduces \TRH{} from 71 to 57. Therefore, we recommend a default \NBO{} of 32 for \defense{} to ensure negligible slowdown even when proactive mitigation is unavailable.

\subsection{Resilience to Performance Attacks}
\begin{figure}[h!]
    \vspace{-0.1in}
    \centering
        \includegraphics[width=0.4\textwidth]{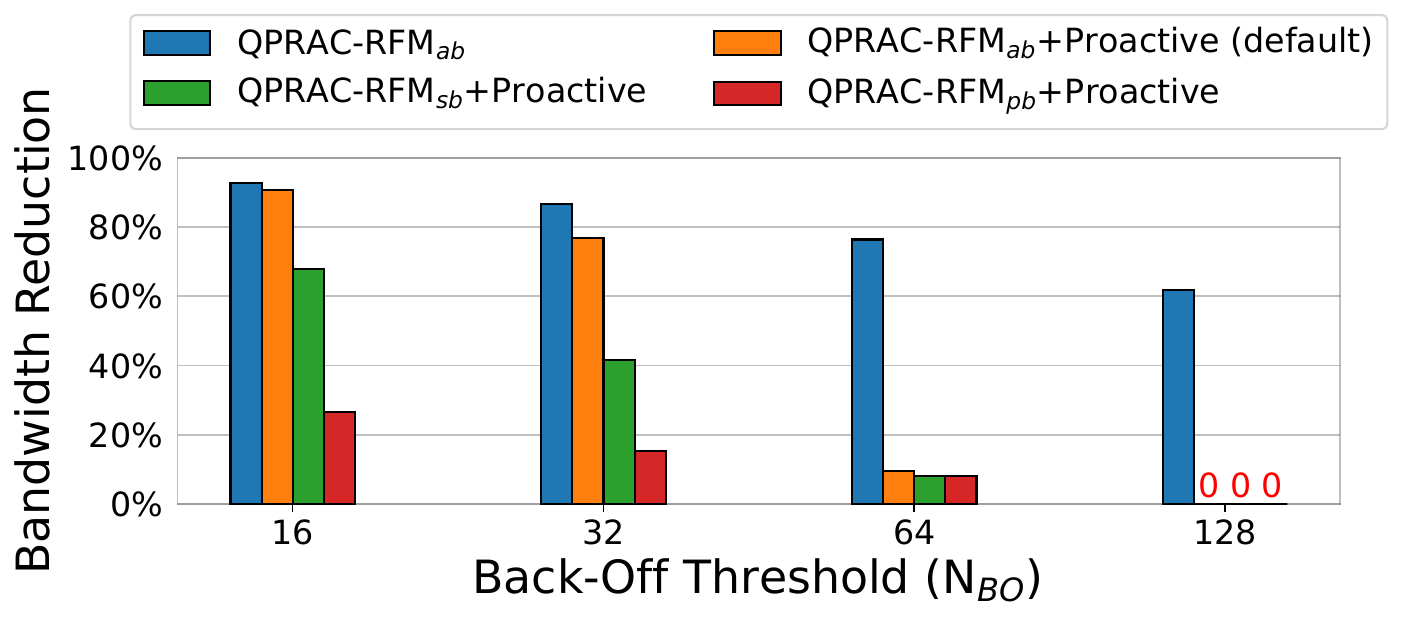}
         \vspace{-0.1in}
    \caption{Maximum DRAM Activation Bandwidth (BW) reduction under attack for \defense{}. \RFMAB{} (all-bank RFM) suffers severe BW degradation, while proactive mitigation addresses this up to \NBO of 64. We note that \RFMSBPB (same-bank/per-bank RFM) is needed at lower \NBO.}
    \label{fig:result:perf_deg}
    \vspace{-0.15in}
\end{figure}

PRAC is vulnerable to performance attacks where an attacker induces a high rate of \ALERT{}s, reducing activation bandwidth for benign workloads. \cref{fig:result:perf_deg} shows the worst-case bandwidth loss for \defense{} at different \NBO{} in a multi-bank attack, where simultaneous hammering across N banks triggers a stream of \ALERT{}s and \RFM{}-induced bandwidth loss.

As per the PRAC specification, All-Bank \RFM{} (\RFMAB) is used on \ALERT{}s (which penalizes all 32 DRAM banks) since the interface does not identify which bank caused the \ALERT{}. 
Consequently, \defense{}-\RFMAB suffers a high loss of activation bandwidth (62\% to 93\%) as \NBO decreases from 128 to 16. With proactive mitigation, \defense{}-\RFMAB{}+Proactive can avoid bandwidth loss at \NBO of 128, and limit it to 10\% at \NBO of 64, as it proactively mitigates rows before they reach activation counts of \NBO. However, at lower \NBO of 32 and 16, this design suffers a bandwidth loss of 77\% and 91\%.

To prevent performance attacks at such \NBO, the PRAC specification can be modified to issue \RFM{}s selectively rather than penalizing all banks with \RFMAB{}. \RFMSB (same-bank), which mitigates one bank in each of the eight bank groups, reduces bandwidth loss to 42\% and 68\% at \NBO of 32 and 16, respectively. \RFMPB (per bank), a new command that mitigates a single bank, further reduces bandwidth loss to 15\% and 27\% at \NBO of 32 and 16. These modifications require changes to the DRAM interface to communicate the mitigation needs to the host and require updates to the DRAM specification.

\subsection{Storage and Energy Overheads}
We evaluate \defense{}'s area overhead, static power, access energy, and circuit latency using the Synopsys Design Compiler with a 45nm process. The area overhead is scaled down to DRAM 10nm process to align with our baseline chip~\cite{DDR5_10nm_chip_size}.

\noindent \textbf{Storage and Latency}:  \defense{} uses a 5-entry priority-based service queue (PSQ) per DRAM bank by default for the mitigation rate of 1 \RFM{} per \ALERT{}. Each entry has a 7-bit activation counter and a 17-bit RowID; together, this requires an SRAM storage of 15 bytes per DRAM bank. Additionally, each DRAM row is accompanied by an activation counter specified by PRAC -- we use 7-bit counters per row.
This incurs 0.038$mm^{2}$ overhead, taking 0.05\% of a single DDR5 chip~\cite{DDR5_10nm_chip_size}. PSQ operations, such as counter increment, comparison, and insertion, take 2.5ns with a 45nm CMOS process and occur in the shadow of Precharge (36ns), introducing no overhead.

\noindent \textbf{Energy}: 
Our synthesis results show the logic for \defense{}'s  PSQ operations (counter increment, comparison, and insertion) consumes an extra 0.23~$\mu\text{J}$ per ACT, just 0.05\% of the activation energy based on the Micron power calculator~\cite{micron:calc}. The QPRAC PSQ, smaller than the TRR trackers in the current DRAM~\cite{hassan2021UTRR}, consumes a static power of 0.38 mW per chip. 

\cref{table:energy_overhead} shows the energy overhead due to the issued mitigations for the different \defense{} designs, as the \RFM{}s per \ALERT{} (PRAC Level) is varied. \defense{} incurs a minimal energy overhead of 1.5\% for all PRAC levels. While \defense{}+Proactive incurs no slowdown, its approach of proactively mitigating the highest activated row in the PSQ of each bank on \textit{every} \REF{} results in a significant energy overhead of 14.6\%. In contrast, the energy-optimized design, \defense{}+Proactive-EA, reduces the energy overhead to 1.9\% by performing proactive mitigations \emph{only} when the highest activated row in the PSQ has an activation count that reaches or exceeds the proactive mitigation threshold (\NPRO{}), while ensuring negligible slowdown. Overall, \defense{} and \defense{}+Proactive-EA incur negligible energy overheads.

\begin{table}[h]
\centering
\vspace{-0.15in}
\caption{Energy Overhead of \defense{}}
\label{table:energy_overhead}
\vspace{-0.1in}
\resizebox{\columnwidth}{!}{%
\begin{tabular}{c|c|c|c}
\toprule
\textbf{PRAC Level} & \textbf{\defense{}} & \textbf{\defense{}+Proactive} & \textbf{\defense{}+Proactive-EA} \\ \midrule
PRAC-1   & 1.2\% &  14.6\% &  1.9\% \\
PRAC-2   & 1.3\% &  14.6\% &  1.9\% \\
PRAC-4   & 1.5\% &  14.6\% &  1.9\% \\
\bottomrule
\end{tabular}%
}
\vspace{-0.1in}
\end{table}

\subsection{Comparison with In-DRAM Mitigations}
\begin{figure}[h!]
    \vspace{-0.1in}
    \centering    \includegraphics[width=0.45\textwidth,height=\paperheight,keepaspectratio]{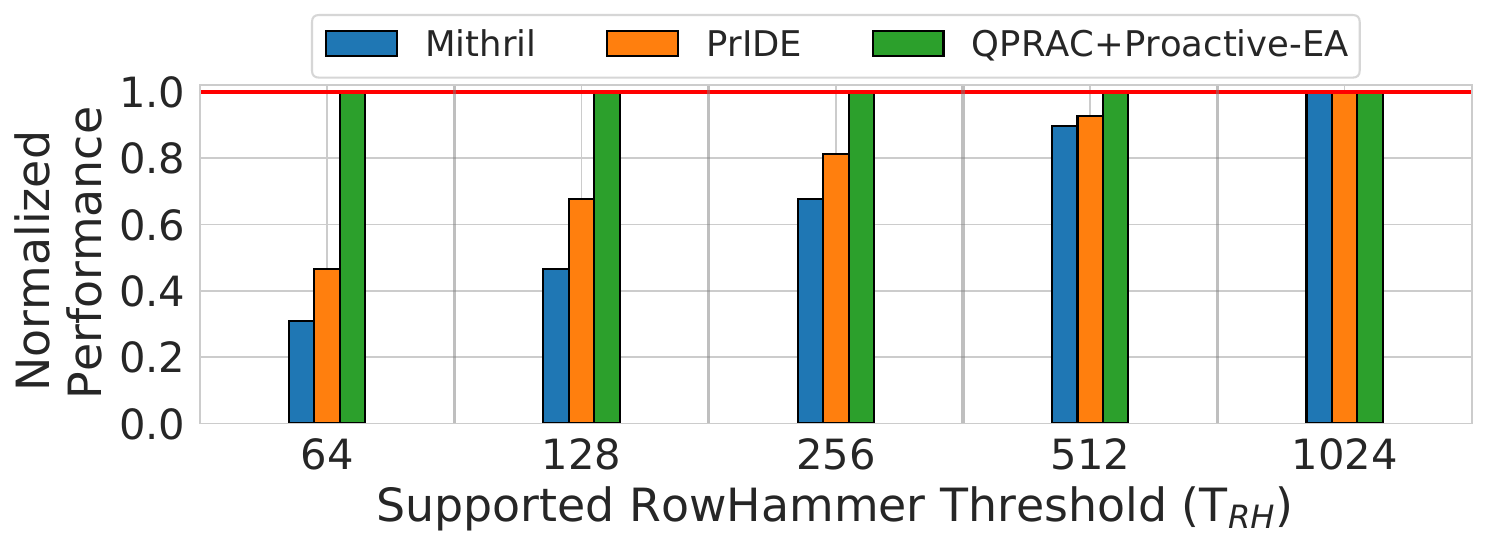}
    \vspace{-0.1in}
    \caption{Normalized performance of \defense{}, Mithril, and PrIDE as the Rowhammer threshold (\TRH{}) varies. Mithril and PrIDE incur significant overhead at ultra-low \TRH{} (\TRH{} $\leq$ 512). Mithril's performance drops from 69\% to 10\% as \TRH{} increases from 64 to 512, while PrIDE shows slowdowns ranging from 54\% to 7\% at the same thresholds. In contrast, \defense{} incurs no performance overhead across all evaluated \TRH{}.
    } 
    \label{fig:result:PrIDE_Mithril}
    \vspace{-0.1in}
\end{figure}

\purple{
\cref{fig:result:PrIDE_Mithril} compares the performance of \defense{} against state-of-the-art in-DRAM Rowhammer mitigations, Mithril~\cite{kim2022mithril} and PrIDE~\cite{jaleel2024pride}, as the Rowhammer threshold (\TRH{}) varies. We assume at most one proactive mitigation per \REF{} and use the DRAM timings for Mithril and PrIDE without PRAC-specific timing increases~\cite{jedec_ddr5_prac}. 
All schemes show minimal slowdown at \TRH{} of 1024.
However, at ultra-low \TRH{} (\TRH{} $\leq$ 512), both Mithril and PrIDE incur significant slowdowns. Mithril's performance drops by 69\%, 54\%, 32\%, and 10\%, at \TRH{} of 64, 128, 256, and 512, while PrIDE shows 54\%, 32\%, 19\%, and 7\% slowdowns at the same \TRH{}. In contrast, \defense{} incurs no slowdown across all evaluated thresholds. 
Additionally, Mithril requires a 5,300-entry CAM/bank, which is impractical, whereas \defense{} only requires a 5-entry CAM/bank, smaller than the TRR trackers in DDR4~\cite{hassan2021UTRR}.
}

\section{Related Work}
\noindent{\bf A. Secure PRAC Designs:} A concurrent work, MOAT\cite{MOAT}, also demonstrated vulnerabilities with PRAC. MOAT showed that in a PRAC implementation like Panopticon, configured for a Rowhammer threshold of 128, tardiness in mitigation using FIFO-based queues can cause activation counts of up to 1150, far beyond the Rowhammer threshold of 128. In comparison, we demonstrate much worse attacks exploiting FIFO-based queues, achieving up to 1300 activations by exploiting \ABOACT{} fundamental to PRAC specification (\textit{Fill+Escape} attack) and up to 30,000 ACTs exploiting the t-bit toggling in Panopticon (\textit{Toggle+Forget} attack).

\begin{figure}[h!]
\vspace{-0.15in}
    \centering
\includegraphics[width=0.45\textwidth, height=\paperheight,keepaspectratio]{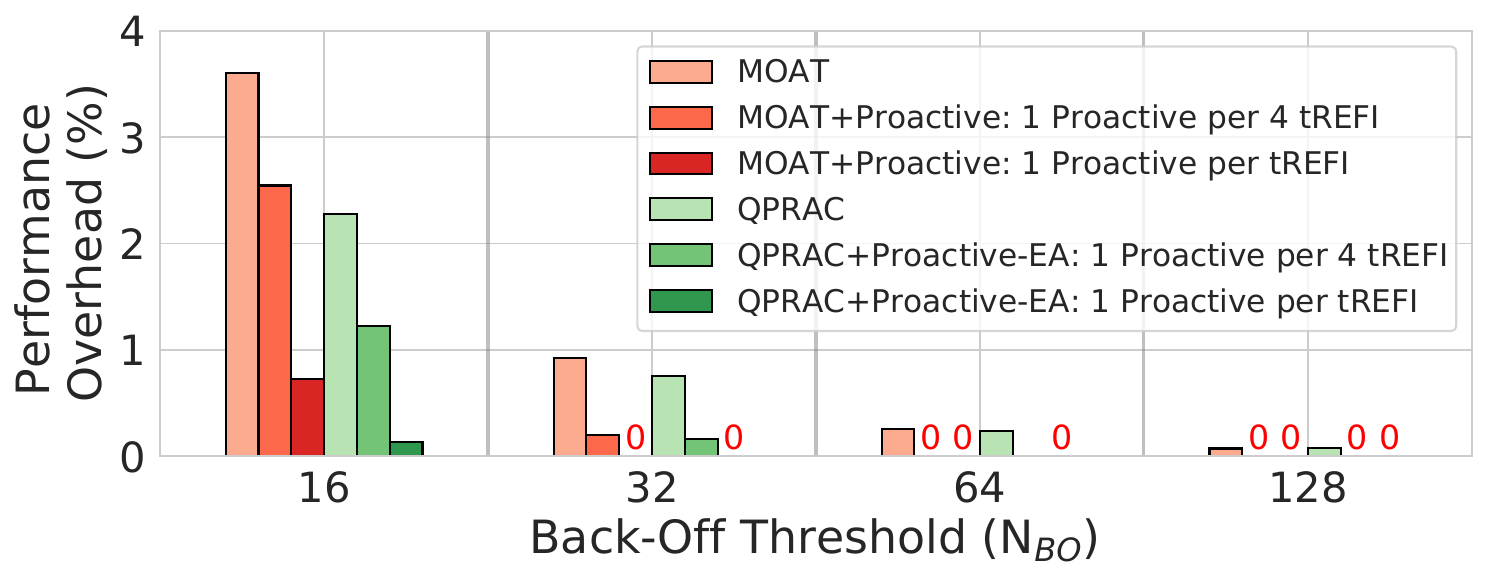}
    \vspace{-0.1in}
    \caption{Slowdown of MOAT\cite{MOAT} and \defense{} as the Back-Off threshold (\NBO{}) varies. At \NBO{} of 32 or more, MOAT and \defense{} show negligible overhead (less than 1\%). At \NBO{} of 16, MOAT and MOAT with 1 proactive mitigation per \TREFI{} incur slowdown of 3.6\% and 0.7\%, while \defense{} and its variant with proactive mitigation have lower slowdowns of 2.3\% and 0.1\%.
    }
    \label{fig:result:moat_vs_qprac_nbo}
    \vspace{-0.10in}
\end{figure}

\cref{fig:result:moat_vs_qprac_nbo} compares the performance of MOAT and \defense{} for PRAC-1 (1 \RFM{} per \ALERT{}) as \NBO{} varies. We assume MOAT uses an enqueuing threshold of \NBO{}$/2$ and a single-entry queue with an additional register, as described in their work\cite{MOAT}. Both \defense{} and MOAT show negligible slowdown ($<$1\%) at \NBO{} of 32 or higher due to the infrequent \ALERT{}s. However, due to its multi-entry queue design, \defense{} outperforms MOAT at lower \NBO{}, benefiting from the opportunistic and proactive mitigations. For example, at \NBO{} of 16, MOAT and its variant with at most one proactive mitigation per 4 \TREFI{} and at most 1 per \TREFI{} shows 3.6\%, 2.5\%, and 0.7\% slowdown. In contrast, \defense{} and its variants with the same proactive mitigation ratios have only a slowdown of 2.3\%, 1.2\%, and 0.1\%, demonstrating better scalability.

\begin{figure}[b!]
    \centering
\includegraphics[width=0.45\textwidth,height=\paperheight,keepaspectratio]{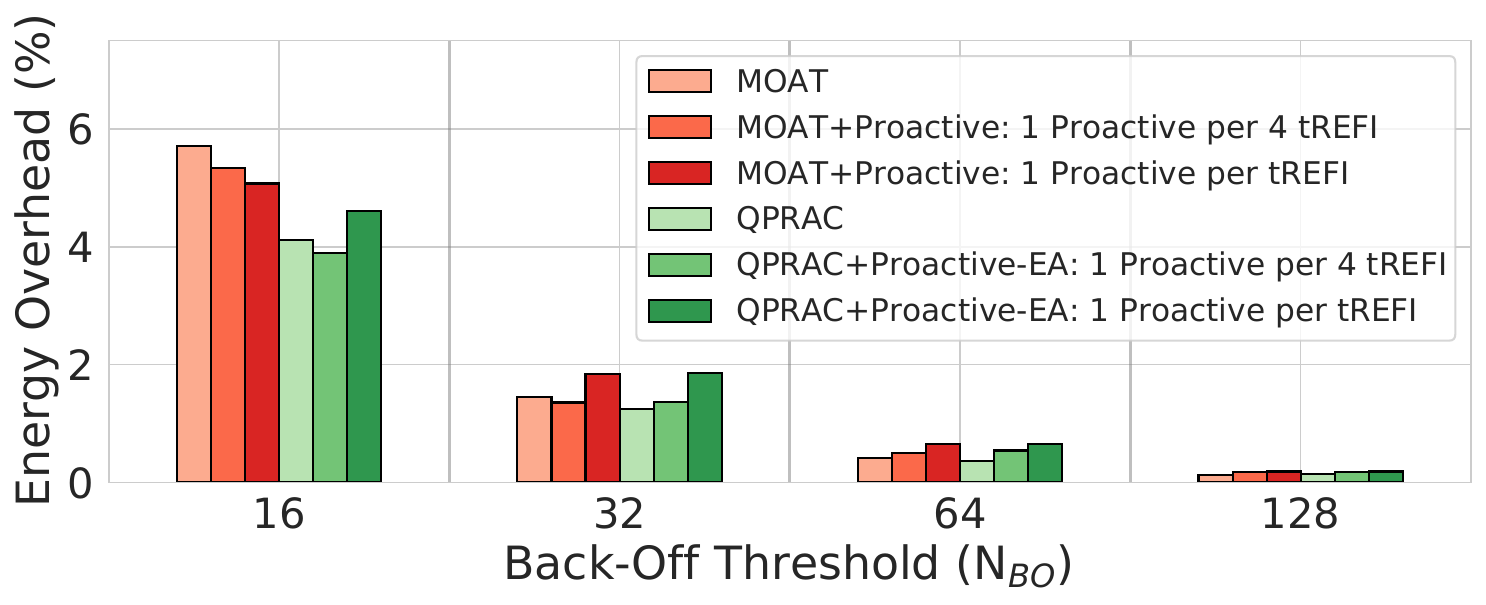}
    \vspace{-0.1in}
    \caption{Energy overhead of MOAT and \defense{} as the Back-Off threshold (\NBO{}) varies. For \NBO{} of 32 or higher, both \defense{} and MOAT show negligible energy overhead (less than 2\%) due to the dual-threshold design (MOAT) and energy-aware proactive mitigation (\defense{}). At \NBO{} of 16, MOAT and MOAT with one proactive mitigation per \TREFI{} incur 5.7\% and 5.1\% overhead, while \defense{} incurs 4.1\% and 4.6\% overhead, respectively.
    }
    \label{fig:result:moat_vs_qprac_energy}
\end{figure}
\Cref{fig:result:moat_vs_qprac_energy} shows the energy overhead of MOAT and \defense{} as \NBO{} varies. At \NBO{} of 32 or higher, both MOAT and \defense{} exhibit negligible energy overhead of less than 2\%, due to the dual-threshold design (MOAT) and energy-aware proactive mitigation (\defense{}).
At \NBO{} of 16, MOAT and MOAT with one proactive mitigation per \TREFI{} incur a 5.7\% and 5.1\% energy overhead. In contrast, \defense{} and its variant with the same proactive mitigation ratio incur a 4.1\% and 4.6\% overhead, due to \defense{}'s reduced execution time.

\noindent{\bf B. One Counter-Per-Row}: PRHT\cite{HynixRH} maintains per-row counters in DRAM to identify aggressor rows, like PRAC\cite{UPRAC, bennett2021panopticon}; however, it reports a 10\% failure rate, suggesting it is insecure. In comparison, \defense{} has deterministic security at \TRH{} as low as 71. Hydra\cite{qureshi2022hydra}, CRA\cite{kim2014architectural}, and START\cite{saxena2024start} store per-row counters in a reserved DRAM region and require extra DRAM accesses by the memory controller to fetch them, causing high worst-case slowdowns at sub-100 \TRH{}. In contrast, we use PRAC-based in-DRAM counters, avoiding such overheads and memory-controller changes.

\noindent{\bf C. Efficient Aggressor Counting}: 
Other works use low-cost in-DRAM trackers using SRAM in DRAM chips, such as CAT\cite{CBT}, TWiCE\cite{lee2019twice}, and Mithril\cite{kim2022mithril} and ProTRR\cite{ProTRR}  (which use Misra-Gries summaries\cite{park2020graphene}). We compare their storage with \defense{} in \cref{table:CompareRelatedWork}. These solutions, feasible at \TRH{} of 4K, requiring KBs of storage, become impractical at \TRH{} below 100 as they require MBs of SRAM. 

Samsung's DSAC\cite{DSAC} and Hynix's PAT\cite{HynixRH} propose probabilistic trackers with less than 20 entries, but DSAC is insecure\cite{PROTEAS}, and PAT has a similar failure rate as conventional trackers. PrIDE\cite{jaleel2024pride}, a probabilistic in-DRAM tracker with four entries per bank, offers security up to \TRH{} of 400 but suffers up to 30\% bandwidth loss at \TRH{} of 250. In comparison, \defense{} requires negligible storage (15 bytes per bank), has deterministic security, and incurs no overhead at sub-100 \TRH{}, surpassing all prior works.

\begin{table}[htb]
  \centering
  \begin{footnotesize}
 \caption{Per-Bank SRAM Overhead of in-DRAM Trackers}
  \label{table:CompareRelatedWork}
  \footnotesize{
  \begin{tabular}{lcc}
    \hline
    \textbf{Name} & \textbf{\TRH{} = 4K} & \textbf{\TRH{} = 100}  \\ \hline
    
    Misra-Gries\cite{park2020graphene} & 42.5 KB & 1700 KB \\ 
    TWiCe\cite{lee2019twice} &  300KB & 12 MB \\
    CAT\cite{CBT} & 196 KB & 7.84 MB\\

    \textbf{\defense{}}&  \textbf{15 bytes}  & \textbf{15 bytes} \\ \hline
  \end{tabular}
  }
  \end{footnotesize}
\end{table}

\noindent {\bf D. Alternative Mitigative Actions}: Memory-controller-based defenses can issue mitigations probabilistically\cite{kim2014architectural, kim2014flipping, hammerfilter, MRLOC, PROHIT}. However, these require knowledge of DRAM neighbors, which may be scrambled within DRAM, or solutions like Directed RFM (DRFM), whose rates are insufficient for sub-100 thresholds. Other methods involve row migration\cite{saileshwar2022RRS, AQUA, SRS, CROW, ShadowHPCA23,saxena2024rubix} or access rate limiting\cite{yauglikcci2021blockhammer}, which introduce high overheads at \TRH{} below 100. Other approaches\cite{HIRA,REGA_SP23} modify the DRAM interface to allow simultaneous refresh or mitigation during activations. In contrast to all these proposals, our solution, \defense{} is in-DRAM, avoids intrusive changes, and maintains low overheads at \TRH{} below 100. 

\smallskip
\noindent{\bf E. Error Detection and Correction}: {SafeGuard}\cite{ali2022safeguard}, {CSI-RH}\cite{csi}, {PT-Guard}\cite{DSN23_PTGuard}, and MUSE\cite{manzhosov2022revisiting} use message authentication codes to detect Rowhammer failures. Failures may also be reduced by scrambling data layouts with ECC\cite{twobirds}; however, uncorrectable bit-flips can still cause data loss. In comparison, \defense{} provides deterministic security against Rowhammer attacks, preventing any possibility of data loss. 

\section{Conclusion}
To enable scalable and secure Rowhammer mitigation, the recent JEDEC DDR5 specification introduces Per Row Activation Counting (PRAC) but provides minimal implementation guidelines. Existing approaches are either insecure or impractical. Our paper addresses these challenges by proposing a PRAC implementation with a scalable priority-based service queue (PSQ) design, ensuring strong security at \TRH below 100. Our solution, \defense{}, leverages opportunistic and proactive mitigations, achieving 0\% slowdown at \TRH of 71 while requiring just 15 bytes per bank.

\section{Acknowledgments}
We thank Stefan Saroiu for sharing insights on PRAC in his keynote at DRAMSec'24 that inspired this work. We also thank Alec Wolman for insightful discussions on Panopticon and PRAC. Special thanks to Kuljit Bains for the feedback and valuable insights on PRAC. We are grateful to SK Hynix, Micron, and Samsung for their feedback, especially for inspiring the energy-optimized design of QPRAC. We also thank Kwangrae Kim for his assistance in the area and power analysis. This work is supported by the Natural Sciences and Engineering Research Council of Canada (NSERC) funding numbers RGPIN-2019-05059 and RGPIN-2023-04796.
\begin{appendices}

\section{Panopticon Blocking \ABOACT{} from Toggling t-bit}
As Panopticon suffers insecurity due to t-bit toggles by \ABOACT{} (\Attackone{} in \cref{sec:pano_attacks}), one way to address this can be to disallow \ABOACT{} from toggling t-bit. 

\begin{figure}[h!]
\vspace{-0.1in}
\centering
\includegraphics[width=\columnwidth,height=\paperheight,keepaspectratio]{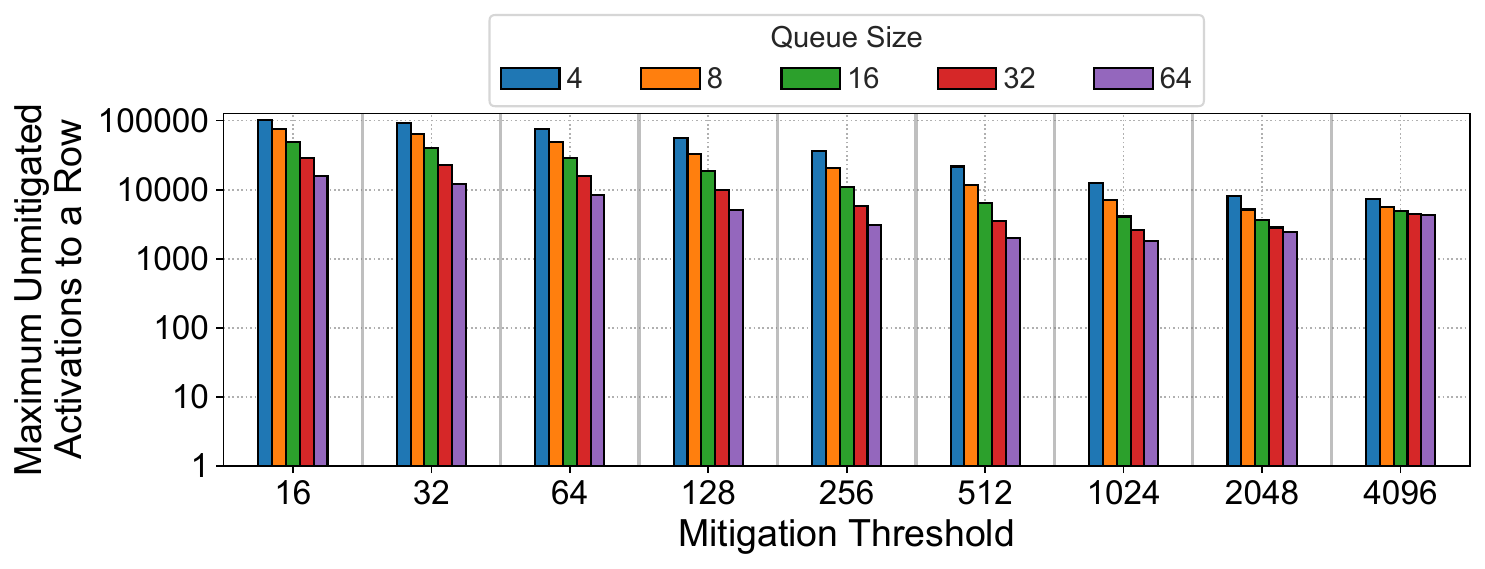}
\vspace{-0.1in}
\caption{The security vulnerability of Panopticon with blocking \ABOACT{} from toggling the threshold bit.}
\label{fig:panopticon+block_ABO_ACT}
\vspace{-0.2in}
\end{figure}

However, now, an adversary can use \ABOACT{} to hammer a target row without mitigation. For a mitigation threshold of M and queue size of Q, an attacker could hammer M - 1 \ACT{}s to Q rows in all 32 banks in parallel and then hammer Q rows $\times$ 1 \ACT{} per bank, get an Alert and 3 \ABOACT{} to hammer the target row, repeating this for 32 banks. \cref{fig:panopticon+block_ABO_ACT} shows the results of this attack. At a minimum, this achieves 1800 \ACT{}s unmitigated for M of 1024.

One could also disallow \ABOACT{} from toggling t-bit \textit{only} for the bank with a full service queue. However, \Attacktwo{} in \cref{sec:attack2} is still possible, where the attacker uses \ABOACT{} while the queue is full to induce Rowhammer, causing 1283 unmitigated activations to a target row. Thus, Panopticon is still insecure below \TRH of 1200.

\section{Artifact Appendix}

%%%%%%%%%%%%%%%%%%%%%%%%%%%%%%%%%%%%%%%%%%%%%%%%%%%%%%%%%%%%%%%%%%%%%
\subsection{Abstract}
This artifact addresses two main aspects of the paper's results: (1) Security analysis of previous FIFO-based PRAC implementations, specifically targeting the Panopticon's t-bit toggle and our priority-based service queue implementation, QPRAC; (2) Performance analysis of QPRAC.

We provide Python scripts for the security analysis to evaluate the security metrics in \cref{eq:2} and \cref{eq:3}. These scripts also regenerate the results presented in Figure~\ref{fig:panopticon_insecure} and Figures~\ref{fig:prac_nonline} to~\ref{fig:prac_trh_proa}.

For the performance analysis, we offer (1) the C++ code for the QPRAC implementation, integrated with Ramulator2~\cite{ramulator2}, and (2) all evaluated workload traces and key evaluation results from Figure~\ref{fig:result:perf} to Figure~\ref{fig:result:PrIDE_Mithril}. Additionally, Bash and Python scripts are provided to automate the collation and plotting of the results.

\subsection{Artifact Check-List (Meta-Information)}
\subsubsection{Security Evaluations}
{\small
\begin{itemize}
  \item {\bf Program: } Python3 programs to evaluate \cref{eq:2} and \cref{eq:3} for different QPRAC configurations. Bash scripts and Python scripts to generate evaluation results and produce plots.
  \item {\bf Run-time environment: } Tested on Ubuntu 20.04 and 22.04 and should run on any Linux Distribution with a valid Python3 interpreter.
  \item {\bf Hardware: } Single Core CPU desktop/laptop suffices.
  \item {\bf Output: } Insecurity of Panopticon: \cref{fig:panopticon_insecure} and QPRAC security evaluation: \cref{fig:prac_nonline} to \cref{fig:prac_trh_proa}
  \item {\bf Experiments: } Instructions to run the experiments and parse the
results are available in the provided README file.
  \item {\bf How much disk space required (approximately)?: } Less than 100MB
  \item {\bf How much time is needed to prepare workflow (approximately)?: } Under 5 minutes to install the dependencies.
  \item {\bf How much time is needed to complete experiments (approximately)?: } $\approx$ 2 hours.
  \item {\bf Publicly available?: } Yes, GitHub: \url{https://github.com/sith-lab/qprac}. 
  \item {\bf Archived (provide DOI)?: } \url{https://doi.org/10.5281/zenodo.14336354}
\end{itemize}
}

\subsubsection{Performance Evaluations}
{\small
\begin{itemize}
    \item {\bf Program:} C++ programs for Ramulator2, including QPRAC implementations, and Bash and Python scripts to run experiments, collate results, and generate plots.
    \item {\bf Compilation:} Tested with g++ version 12.4.0; should also compile with any C++20-compliant compiler.
    \item {\bf Run-time environment:} We suggest using a Linux distribution compatible with g++-10 or newer for the performance evaluations. For example, Ubuntu 22.04 or later is recommended if you prefer Ubuntu. This artifact has been tested on Ubuntu 22.04 and Rocky Linux 9.4.
    \item {\bf Hardware:} We recommend using a CPU with at least 40 cores and 128GB or more of memory.
    \item {\bf Metrics:} Weighted Speedup.
    \item {\bf Output:} QPRAC performance results: Figures~\ref{fig:result:perf} to Figure~\ref{fig:result:PrIDE_Mithril}.
    \item {\bf Experiments:} Instructions for running experiments and parsing results are available in the provided README file.
  \item {\bf How much disk space required (approximately)?: } $\approx$ 10GB.
  \item {\bf How much time is needed to prepare workflow (approximately)?:} Under 10 minutes to install the dependencies and download the traces.
  \item {\bf How much time is needed to complete experiments (approximately)?: }     \begin{itemize}
        \item $\approx$ 16 hours (on a cluster with 1000 cores) and $\approx$ 1 day (on an Intel Xeon CPU with 40 cores and 128GB memory) for the main results (Figures~\ref{fig:result:perf} and~\ref{fig:result:num_alert}).
        \item $\approx$ 2 days (on a cluster with 1000 cores) and $\approx$ 1 week (on an Intel Xeon CPU with 40 cores and 128GB memory) for all experiments (Figures~\ref{fig:result:perf} to~\ref{fig:result:PrIDE_Mithril}).
    \end{itemize}
  \item {\bf Publicly available?:} Yes. Traces:  \url{https://zenodo.org/records/14607144}. GitHub: \url{https://github.com/sith-lab/qprac}. 
  \item {\bf Archived (provide DOI)?: }\url{https://doi.org/10.5281/zenodo.14336354}
\end{itemize}
}

%%%%%%%%%%%%%%%%%%%%%%%%%%%%%%%%%%%%%%%%%%%%%%%%%%%%%%%%%%%%%%%%%%%%%
\subsection{Description}

\subsubsection{How to access}
The artifact is available at \url{https://github.com/sith-lab/qprac}.
\subsubsection{Hardware dependencies}
\begin{itemize}
    \item \textbf{Security Evaluation:} A single-core CPU desktop/laptop will allow to perform security analysis within $\approx$ 2 hours.
    \item \textbf{Performance Evaluation: }We strongly recommend using Slurm with a cluster capable of running bulk experiments to accelerate evaluations. If you opt for a personal server, we advise using a CPU with at least 40 cores and 128GB or more memory.
\end{itemize}

\subsubsection{Software dependencies}
\begin{itemize}
    \item \textbf{Security Evaluation:}
        \begin{itemize}
            \item Python3 (Tested on V3.11.5).
            \item Python3 Package matplotlib (v.3.4.0 or higher is required) for plotting.
        \end{itemize}
    \item \textbf{Performance Evaluation:}
        \begin{itemize}
        \item g++ with C++20 support (tested with version 12.4.0).
        \item Python3 (recommended: version 3.10 or above).
    \end{itemize}
\end{itemize}

\subsubsection{Traces}

We use 57 traces from SPEC2006, SPEC2017, TPC, Hadoop, MediaBench, and YCSB, available for download at \url{https://zenodo.org/records/14607144}.

%%%%%%%%%%%%%%%%%%%%%%%%%%%%%%%%%%%%%%%%%%%%%%%%%%%%%%%%%%%%%%%%%%%%%
\subsection{Installation and Experiment Workflow}
First, clone the GitHub repository: 
\begin{tcolorbox}[colback=gray!10, colframe=black, boxrule=0.5pt, sharp corners, width=\linewidth, arc=0mm, top=1pt, bottom=1pt]
\textbf{\$ git clone https://github.com/sith-lab/qprac.git}
\end{tcolorbox}

\subsubsection{Security Evaluation} 
No additional setup is required if dependencies are satisfied.

To start the experiment:
\begin{tcolorbox}[colback=gray!10, colframe=black, boxrule=0.5pt, sharp corners, width=\linewidth, arc=0mm, top=1pt, bottom=1pt]
\textbf{\$ cd qprac/security\_analysis}\\
\textbf{\$ bash ./run\_artifact}
\end{tcolorbox}

To use provided sample data and not regenerate results:
\begin{tcolorbox}[colback=gray!10, colframe=black, boxrule=0.5pt, sharp corners, width=\linewidth, arc=0mm, top=1pt, bottom=1pt]
\textbf{\$ cd qprac/security\_analysis}\\
\textbf{\$ bash ./run\_artifact --use-sample}
\end{tcolorbox}

\subsubsection{Performance Evaluation}
\noindent 1. Configure the following parameters in \texttt{perf\_analysis/run\_artifact.sh}: 
\begin{itemize}
    \item Using Slurm:
        \begin{itemize}
            \item SLRUM\_PART\_NAME: Partition name for Slurm jobs. 
            \item SLRUM\_PART\_DEF\_MEM: Default memory size for jobs (recommended: $\geq$ 4GB). 
            \item SLRUM\_PART\_BIG\_MEM: Memory size for jobs that require large memory (recommended: $\geq$ 12GB).
            \item MAX\_SLRUM\_JOBS: Maximum number of Slurm jobs to submit.
        \end{itemize}
    \item Using a Personal Server:
        \begin{itemize}
            \item PERSONAL\_RUN\_THREADS: Number of parallel threads to use for simulations.
        \end{itemize}    
\end{itemize}

\noindent 2. Run the following commands to install dependencies, build Ramulator2, and execute simulations. If using a personal server with limited resources (e.g., less than 256GB memory or 40 cores), we recommend running only the main experiments first to avoid long execution times and then running the remaining experiments.

\noindent 2.1. Running main experiments (Figures~\ref{fig:result:perf} and~\ref{fig:result:num_alert}):
\begin{itemize}
    \item Using Slurm:
    \begin{tcolorbox}[colback=gray!10, colframe=black, boxrule=0.5pt, sharp corners, width=\linewidth, arc=0mm, top=1pt, bottom=1pt, halign=left]
    \textbf{\$ cd qprac/perf\_analysis\\ 
    \$ ./run\_artifact.sh \texttt{--}method slurm \texttt{--}artifact main}
    \end{tcolorbox}
    \item Using a Personal Server:
    \begin{tcolorbox}[colback=gray!10, colframe=black, boxrule=0.5pt, sharp corners, width=\linewidth, arc=0mm, top=1pt, bottom=1pt, halign=left]
    \textbf{\$ cd qprac/perf\_analysis\\ 
    \$ ./run\_artifact.sh \texttt{--}method personal \texttt{--}artifact main}
    \end{tcolorbox}
\end{itemize}
\noindent 2.2. Running all experiments (Figures~\ref{fig:result:perf} to~\ref{fig:result:PrIDE_Mithril}):
\begin{itemize}
    \item Using Slurm:
    \begin{tcolorbox}[colback=gray!10, colframe=black, boxrule=0.5pt, sharp corners, width=\linewidth, arc=0mm, top=1pt, bottom=1pt, halign=left]
    \textbf{\$ cd qprac/perf\_analysis\\ 
    \$ ./run\_artifact.sh \texttt{--}method slurm \texttt{--}artifact all}
    \end{tcolorbox}
    \item Using a Personal Server:
    \begin{tcolorbox}[colback=gray!10, colframe=black, boxrule=0.5pt, sharp corners, width=\linewidth, arc=0mm, top=1pt, bottom=1pt, halign=left]
    \textbf{\$ cd qprac/perf\_analysis\\ 
    \$ ./run\_artifact.sh \texttt{--}method personal \texttt{--}artifact all}
    \end{tcolorbox}
\end{itemize}
%%%%%%%%%%%%%%%%%%%%%%%%%%%%%%%%%%%%%%%%%%%%%%%%%%%%%%%%%%%%%%%%%%%%%
\subsection{Evaluation and Expected Results}
\subsubsection{Security Evaluation}
The artifact provides the following scripts in the qprac/security\_analysis/analysis\_scripts directory: \texttt{equation2.py}, \texttt{equation2\_pro.py}, \texttt{equation3.py}, \texttt{equation3\_pro.py}, and \texttt{tbit\_attack.py}. These scripts allow for the collation of results, with commands for generating the results for \cref{fig:panopticon_insecure}, \cref{eq:2}, and \cref{eq:3} provided in the \texttt{run\_artifact.sh} script and the README file. After running \texttt{run\_artifact.sh}, the t-bit attack results, along with the results for \cref{eq:2} and \cref{eq:3} in both QPRAC and QPRAC + Proactive, will be available as \texttt{tbit\_attack.txt}, \texttt{PRAC1-4.txt}, \texttt{PRAC1-4\_PRO.txt}, \texttt{R1.txt}, and \texttt{R1\_PRO.txt} files in the qprac/security\_analysis directory. Additionally, the regenerated figures, including \cref{fig:panopticon_insecure} and \cref{fig:prac_nonline} to \cref{fig:prac_trh_proa}, can be found as \texttt{figure\#.pdf} files in the corresponding qprac/security\_analysis/figure\# folders. Sample data required to generate each figure is provided in the qprac/security\_analysis/figure\#/sample\_data folders.

\subsubsection{Performance Evaluation}
After completing the experiments using \texttt{run\_artifact.sh}, the results and plots can be regenerated with the provided scripts. Specifically, the artifact includes the \texttt{plot\_main\_figures.sh} and \texttt{plot\_all\_figures.sh} files in the qprac/perf\_analysis directory. These scripts collate the results (obtained as CSV files in qprac/perf\_analysis/results/csvs) and generate the plots (obtained as PDF files in qprac/perf\_analysis/results/plots). The \texttt{plot\_main\_figures.sh} script regenerates Figures~\ref{fig:result:perf} and~\ref{fig:result:num_alert}, while the \texttt{plot\_all\_figures.sh} script generates Figures~\ref{fig:result:perf} to~\ref{fig:result:PrIDE_Mithril}. Additionally, we provide scripts to collate results (\texttt{generate\_csv\_fig\#.py}) and generate plots (\texttt{plot\_fig\#.py} or \texttt{plot.ipynb}) for each experiment in the qprac/perf\_analysis/plot\_scripts directory. Sample result files and plots are available in the qprac/perf\_analysis/results/sample\_results directory.

%%%%%%%%%%%%%%%%%%%%%%%%%%%%%%%%%%%%%%%%%%%%%%%%%%%%%%%%%%%%%%%%%%%%%
\subsection{Experiment Customization: Performance Evaluation}
We offer easy configuration options for the following parameters: 
1) the evaluated QPRAC mechanisms, 
2) the evaluated Back-Off Thresholds (\NBO{}), 
3) the tested PRAC levels (number of RFMs per alert), and 
4) the simulation duration (minimum number of instructions per core during experiments).

These parameters can be customized in the \texttt{qprac/perf\_analysis/sim\_scripts/run\_config\newline
\_fig\#.py} files:
\begin{itemize}
    \item Mitigation list, PSQ sizes, and proactive mitigation frequencies: \texttt{mitigation\_list}, \texttt{psq\_sizes}, and \texttt{targeted\_ref\_ratios}.
    \item Back-Off Thresholds: \texttt{NBO\_lists}.
    \item PRAC levels: \texttt{PRAC\_levels}.
    \item Simulation duration: \texttt{NUM\_EXPECTED\_INSTS}.
\end{itemize}

%%%%%%%%%%%%%%%%%%%%%%%%%%%%%%%%%%%%%%%%%%%%%%%%%%%%%%%%%%%%%%%%%%%%%
\subsection{Methodology}

Submission, reviewing and badging methodology:

\begin{itemize}
  \item \url{https://www.acm.org/publications/policies/artifact-review-and-badging-current}
  \item \url{https://cTuning.org/ae}
\end{itemize}

\end{appendices}

%%%%%%% -- PAPER CONTENT ENDS -- %%%%%%%%

%%%%%%%%% -- BIB STYLE AND FILE -- %%%%%%%%

\bibliographystyle{IEEEtranS}
\balance
\bibliography{main}
%%%%%%%%%%%%%%%%%%%%%%%%%%%%%%%%%%%%

\end{document}